\documentclass[10pt, prb, aps, amssymb, twocolumn, superscriptaddress, notitlepage, longbibliography]{revtex4-2}

\usepackage{amsmath}
\usepackage{bbold}
\usepackage{graphicx}
\usepackage{color}
\usepackage{bbm}
\usepackage[normalem]{ulem}
\usepackage[colorlinks]{hyperref}
\usepackage{dsfont}
\usepackage{physics}
\usepackage{enumitem}
\usepackage{float}
\usepackage[english]{babel}

\newcommand{\pd}{{\phantom{\dag}}}

\begin{document}
\title{Tunable Chern superconductivity of PtBi$_2$ in slab geometry}

\author{Luca Ketmaier}
\affiliation{Leibniz Institute for Solid State and Materials Research, IFW Dresden, Helmholtzstrasse 20, 01069 Dresden, Germany}
\affiliation{W\"urzburg-Dresden Cluster of Excellence ctd.qmat}

\author{Jeroen van den Brink}
\affiliation{Leibniz Institute for Solid State and Materials Research, IFW Dresden, Helmholtzstrasse 20, 01069 Dresden, Germany}
\affiliation{W\"urzburg-Dresden Cluster of Excellence ctd.qmat}
\affiliation{Institute for Theoretical Physics, Technische Universit\"at Dresden, 01062 Dresden, Germany}

\author{Ion Cosma Fulga}
\affiliation{Leibniz Institute for Solid State and Materials Research, IFW Dresden, Helmholtzstrasse 20, 01069 Dresden, Germany}
\affiliation{W\"urzburg-Dresden Cluster of Excellence ctd.qmat}

\begin{abstract}
Recent experiments indicate that PtBi$_2$ is a Weyl semimetal whose surfaces become topological superconductors with \textit{i}-wave gap symmetry at low temperature. 
The bulk hosts 12 Weyl cones while in the superconducting state each of the two surfaces hosts in addition 6 Majorana cones. 
We study a simplified model in a slab geometry, where the finite thickness gaps out the bulk Weyl cones and leaves the surface Majorana cones as the low-energy degrees of freedom. 
We identify two competing mechanisms that gap these cones: inter-cone hybridization opens a trivial gap, whereas a Zeeman field opens a nontrivial one. 
In the nontrivial regime, chiral Majorana edge modes connect the gapped cones. We determine their wavefunction profile and relate it to the momentum-space distribution of the cones. Mapping out the phase diagram, we find both trivial and nontrivial regions in the quasi-two-dimensional limit.
\end{abstract}

\maketitle

\section{Introduction}
\label{sec:intro}

Over the past three decades, topological superconductivity has formed an increasingly active research direction \cite{Alicea_2012, Beenakker_2013, Sato_2016, Sato_2017, Sharma_2022}.
The goal is to realize a superconducting state with nontrivial topology, as signaled by the nonzero value of some characteristic topological invariant \cite{Kitaev_2001, Schnyder_2008, Qi_2010}.
In many cases, the bandstructure of a topological superconductor is thought of as being analogous to that of a topological insulator \cite{Qi_2011}.
The bulk is fully gapped, whereas robust, mid-gap modes are localized on the boundary: so-called Majorana bound states.
It was recognized that such states could not exist in the simplest (conventional) \textit{s}-wave superconductors \cite{Sato_2016} unless external fields or heterostructures are employed \cite{Sato_2010, Sau_2010, Lutchyn_2010, Mourik_2012},
as the even parity pairing symmetry in such systems implies a topologically trivial band structure \cite{Alicea_2012}.

Not all forms of topological superconductivity require a gapped bulk, however.
Instead, superconductors with nodes in their gap structure \cite{Won_2004}, which are thus unconventional, can have non-trivial topological invariants associated to these nodes.
If the node is formed from a nondegenerate Andreev bound state, it is referred to as a Majorana cone, being characterized by a winding number \cite{Wen_2002, Beri_2010} equal to $\pm 1$.
On the boundary, the projections of Majorana cones with opposite winding numbers must then be connected by mid-gap, dispersionless Majorana edge bands \cite{Sato_2011, Schnyder_2015}.
This scenario is analogous to that of graphene, whose two-dimensional (2D) bulk Dirac cones lead to the formation of zig-zag edge modes at the Fermi level \cite{Nakada_1996, Ryu_2002, Delplace_2011}.

In this context, PtBi$_2$ has emerged as a promising candidate for realizing intrinsic topological superconductivity \cite{Veyrat_2023, Kuibarov_2024, Schimmel_2024, Changdar_2025, Moreno_2026}, i.e.~a nontrivial state which appears without the need for external fields or heterostructures.
The material is a Weyl semimetal \cite{Burkov_2011, Armitage_2018} hosting $12$ Weyl cones close to the Fermi level, which are connected pairwise by surface states known as Fermi arcs \cite{Shipunov_2020, Vocaturo_2024, arxiv.2607.01947, arxiv.2607.10937, arxiv.2607.26804}.
When brought to temperatures below 10 K \cite{Kuibarov_2024, Moreno_2026}, its surfaces become superconducting, whereas bulk superconductivity has been reported below $0.5$ K \cite{Zhang_2025}. 
Experimental evidence suggests that it is the nondegenerate Fermi arc surface states which acquire a nonzero superconducting gap \cite{Kuibarov_2024}, and that there exists a node at the center of each arc, at which the gap vanishes.
Given the symmetries of the material, the presence of nodes would imply that the surface of PtBi$_2$ forms an \textit{i}-wave superconductor \cite{Changdar_2025, Stewart_2017}, a scenario which 
is attracting considerable theoretical interest \cite{Buccheri_2026, Maeland_2026, Dsouza_2026, Waje_2025}. 
Further, since the Fermi arcs themselves are nondegenerate, each superconducting node is by definition a Majorana cone, leading to the formation of mid-gap Majorana flat bands at the ``boundaries of the surface,'' i.e.~the hinges or step-edges of the material.

Here, we consider how the topological surface superconductivity of PtBi$_2$ can be tuned by varying the thickness of finite-size slabs to access different topological phases.
To this end, we use a minimal toy model \cite{Vocaturo_2024} which captures only the symmetries of the material and the essential topological information: the Weyl cones and Fermi arcs.
On the one hand, this means that we neglect all other bulk Fermi pockets, which may obscure and hybridize with the topological states in the real material.
On the other hand, the topological information alone is sufficient to make predictions as to the behavior expected in PtBi$_2$ experiments, while at the same time enabling the numerical simulation of system sizes that would be out of reach in more realistic models.

Since we are interested specifically in the topological superconductivity, we consider a thin slab geometry, such that the normal metal states (the Weyl cones) become gapped due to finite-size effects.
In the resulting quasi-2D system, Majorana cones can undergo different gapping mechanisms leading to different topological invariants, similar to how the bandstructure of the honeycomb lattice can be gapped out  by a Semenoff mass \cite{Semenoff_1984} or by a Haldane mass \cite{Haldane_1988}.
A trivial mass term is generated when Majorana cones on the two different surfaces of the thin slab couple to each other.
Thus, it depends both on the slab thickness as well as the momentum-space position of the Majorana cones in the slab Brillouin zone (BZ), which may be controllable by doping.
In contrast, nontrivial gaps are generated by a Zeeman field applied perpendicular to the slab surfaces, e.g.~as would occur in proximity to a ferromagnet \cite{Men_shov_2013, Chang_2013}.
In this case, the quasi-2D system hosts 6 chiral Majorana modes \cite{Read_2000}, propagating unidirectionally on the side edges of the thin slab.
Further, the thickness of these chiral modes in the direction perpendicular to the edge depends on how the Majorana cones are projected onto the edge BZ.
Thus, it varies with the edge orientation.

This manuscript is organized as follows:
In Sec.~\ref{sec:model} we describe the toy-model Hamiltonian, and in Sec.~\ref{sec:cone_gaps} we show the different effects of the mentioned manipulations.
Finite size effects gap out Weyl cones while shifts in surface potential and applied Zeeman field gap out the Majorana cones.
We determine the topological phase diagram that results from the competition between the two different gapping mechanisms of the Majorana cones in Sec.~\ref{sec:chern_sc}, and examine the chiral edge modes in Sec.~\ref{sec:chiral_edge}.
Finally we summarize our results and discuss implications for future experiments in Sec.~\ref{sec:conc}.

\section{Toy model}
\label{sec:model}

We use the toy model introduced in Ref.~\cite{Vocaturo_2024} and subsequently used in Refs.~\cite{Maeland_2026, Dsouza_2026} to investigate the potential pairing mechanisms leading to nodal superconductivity.
It is a 4-band model in which the unit cell consists of two spinful orbitals, with Pauli matrices $\sigma_j$ and $\tau_j$ parameterizing the spin and orbital degrees of freedom, respectively.
The bulk Hamiltonian is composed of three terms:
\begin{equation}
    H_\text{bulk} = h_0 + \alpha h_1 + \gamma \tau_x \sigma_0,
    \label{eq:ham_general}
\end{equation}
the first is the parent, spin-diagonal Hamiltonian, the second is the spin-mixing term, the third is the inversion symmetry breaking term.
These have the following form:
\begin{align}
    h_0(\mathbf{k}) = &\big[\mu - t \cos(k_1) - t \cos(k_2) - t \cos(k_1 + k_2)
    \nonumber \\
    &+ \beta \cos(k_3) \big]\Gamma_1 + \beta \sin(k_3) \Gamma_3  \nonumber \\
    &+ \lambda \big[\sin(k_1) + \sin(k_2) - \sin(k_1 + k_2) \big] \Gamma_3
    \label{eq:parent_ham}
\end{align}
where $\mathbf{k}$ is the three-dimensional (3D) momentum vector, $k_i = \mathbf{a}_i \cdot \mathbf{k}$, and $\mathbf{a}_i$ are the Bravais vectors of the lattice: $\mathbf{a}_1 = (0,1,0)$, $\mathbf{a}_2 = (\sqrt{3}/2,-1/2,0)$, $\mathbf{a}_3 = (0,0,1)$, in units of the lattice constant. 
Thus, the 3D model consists of triangular lattices stacked on top of each other in the $z$ direction.

The $\Gamma$ matrices are defined as
\begin{align}
    \Gamma_1 = \tau_z \sigma_0 \ \ ; \ \ \Gamma_2 = \tau_x\sigma_x \ \ 
    ; \ \ \Gamma_3 = \tau_y \sigma_0,
\end{align}
and the spin-mixing term has the form
\begin{align}
    h_1(\mathbf{k}) =& \big[1 - \cos(k_3)\big] \big[\sin(k_1)\Gamma_2\nonumber\\
    &+ \sin(k_2)\Gamma_{2,1} - \sin(k_1 + k_2)\Gamma_{2,2} \big],
    \label{eq:SOC_ham}
\end{align}
where $\Gamma_{2,j} = C_3^j \Gamma_2 C_3^{-j}$ and $C_3$ encodes the rotation by $2\pi/3$ along the $z$ axis,
$$
C_3 = \tau_0 \exp(-i \frac{\pi}{3}\sigma_z).
$$
The model obeys time-reversal symmetry, a threefold rotation symmetry along the $z$ direction, and three reflection symmetries characterized by vertical mirror planes, as discussed in Ref.~\cite{Vocaturo_2024}.
Unless otherwise specified, the model parameters throughout the paper are chosen as: in-plane, intra-orbital hopping $t = 1$ (our unit of energy), chemical potential $\mu = 1.7$, inter-orbital hopping $\lambda = 3$, vertical hopping $\beta = -0.75$, spin-orbit coupling strength $\alpha = 0.75$, and inversion symmetry breaking strength $\gamma = 0.5$.
For these parameter values, the Hamiltonian Eq.~\eqref{eq:ham_general} hosts 12 bulk Weyl cones.

We consider a slab geometry, 
having a finite number $N_z$ of unit cells (layers) along the $z$ direction while preserving the original translation symmetry for the other two directions.
The bulk Weyl cones are now connected pairwise by surface Fermi arcs, such that there appear six arcs on the top surface and six on the bottom surface. 
To study surface superconductivity, we perform a Bogoliubov-de Gennes (BdG) transformation on this slab Hamiltonian, writing
\begin{align}
    \mathcal{H} = \begin{pmatrix}
        H^\pd_\text{slab} & \Delta \\ 
        \Delta^\dagger & -H_\text{slab}^T
    \end{pmatrix}.
    \label{eq:BdG_ham}
\end{align}
As usual, the hole block has opposite signs of the momenta.
The pairing matrix $\Delta$ depends on two momenta and one real-space variable.
The pairing term takes the following form: 
\begin{align}
    \Delta(k_1, k_2, z)& =i  \tau_0 \sigma_y f_1(z) \times\\ 
    &\big[ \sin(k_1)\sigma_y + \sin(k_2)
    \sigma_{y,1}\nonumber\\
        &- \sin(k_1 + k_2)\sigma_{y,2} + \sin(k_1 + 2k_2)\sigma_x \nonumber\\
        &- \sin(2k_1 + k_2)\sigma_{x,1} 
        + \sin(k_1 - k_2)\sigma_{x,2} \big] \nonumber,
        \label{eq:Delta_mat}
\end{align}
obtained by imposing the $i$-wave structure described in Refs.~\cite{Changdar_2025, Timm_2021}.
Here, $z = 1,2,\ldots,N_z$ is the layer index and the prefactor $f_1(z)$ is either $1$ or $0$ depending on whether the layer being considered is metallic or superconducting. 
Throughout this work we will consider two superconducting layers on the top and bottom of the slab:
\begin{equation}
    f_1(z) = \begin{cases} 
    1 \ \ \text{if} \ z \le 2 \ \text{or} \  z \ge N_z-1\\
    0 \ \ \text{otherwise}\end{cases}.
\end{equation}
The $\sigma_{\alpha, j}$, $(\alpha=x,y)$ matrices are defined as
\begin{equation}
    \sigma_{\alpha, j} = C_3^j \sigma_{\alpha}C_3^{-j}.
\end{equation}

Finally, motivated by the physical difference between the two surfaces of PtBi$_2$ which have been labeled as ``kagome type'' and ``decorated-honeycomb type,'' \cite{Shipunov_2020, Vocaturo_2024},
we introduce a spatial perturbation in the chemical potential $\mu$ of Eq.~\eqref{eq:parent_ham}, $\mu \rightarrow \mu(z)$, with
\begin{equation}
    \mu(z) = \begin{cases}
        \mu  + \mu_\text{surf} \ \ \text{if} \ \ z \ge N_z-1\\
        \mu   \ \ \text{if} \ N_z-1< z < 2\\
        \mu  - \mu_\text{surf} \ \ \text{if} \ \ z \le 2\\
    \end{cases}
    \label{eq:mu_term}
\end{equation}
where we will take the magnitude of the perturbation to be $\mu_\text{surf} = 0.3$ unless otherwise specified. 
The effect of this antisymmetric surface term is to separate the Fermi arcs and Majorana cones in momentum space.

\begin{figure}[tb]
\centering
\includegraphics[width=0.9\columnwidth]{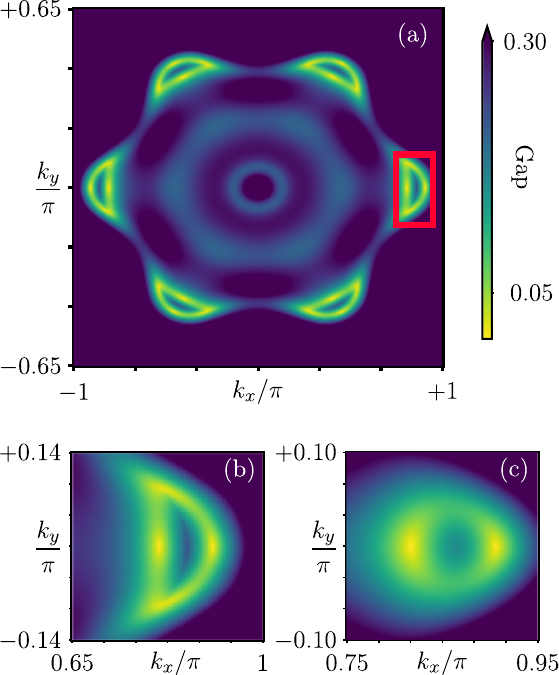}
\caption{
Panel (a): Gap as a function of the two dimensionless momenta for a slab thickness $N_z=50$.
We note that here and throughout the rest of this manuscript, the ``gap'' is defined as the absolute value of the eigenenergy closest to zero for a particular set of momenta and parameters.
The inset in panel (b) shows a zoom in of two Weyl cones and two Majorana cones, corresponding to the red box of panel (a). 
In panel (c), the slab thickness has been reduced to $N_z=4$. 
The Weyl cones have been gapped out, whereas Majorana cones remain clearly visible.}
\label{fig:BZ_gap}
\end{figure}

In the full BdG Hamiltonian Eq.~\eqref{eq:BdG_ham}, Weyl cones remain gapless since the pairing term vanishes in the bulk.
On the two surfaces of the slab, however, each Fermi arc becomes gapped and a node forms at the arc center.
Thus, there are a total of six Majorana cones on the top surface and six other Majorana cones on the bottom surface.
We show this behavior numerically by constructing the slab Hamiltonian with the help of the kwant package \cite{Groth_2014}.
In Fig.~\ref{fig:BZ_gap}(a) we plot the gap (i.e.~the magnitude of the eigenvalue closest to $E=0$) as a function of the dimensionless, Cartesian momenta, $k_x = (k_1+2k_2)/\sqrt{3}$ and $k_y=k_1$, for a slab consisting of $N_z=50$ unit cells.
A total of 12 Weyl cones, indicated by bright colors, are positioned symmetrically in the slab BZ.
Each pair of Weyl cones is connected by two Fermi arcs, one on the top surface and one on the bottom surface.
Finally, the color gradient along the Fermi arcs [visible in the zoomed-in inset of Fig.~\ref{fig:BZ_gap}(b)] shows that each arc is gapped and hosts a Majorana cone at its center.

\section{Gapping out bulk and surface modes}
\label{sec:cone_gaps}

\begin{figure}[tb]
\centering
\includegraphics[width=\columnwidth]{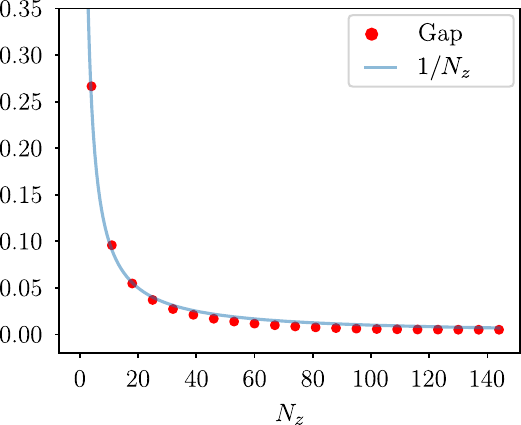}
\caption{Weyl cone gap as a function of slab thickness $N_z$, in units of the intra-orbital hopping $t$. 
The red dots show gap value at $(k_x, k_y) = (0.81 ,  0.09)\pi$, chosen to track one particular Weyl cone.
The blue line is the function $1/N_z$.
}
\label{fig:Weyl_gapping}
\end{figure}

The Weyl cones are formed from 3D extended bulk states, and as such are sensitive to the thickness of the slab.
We take advantage of this fact in order to gap out the Weyl cones by reducing the slab thickness \cite{Wu_2020}.
As shown in Fig.~\ref{fig:Weyl_gapping}, reducing $N_z$ leads to a splitting of the Weyl bands that scales as the inverse thickness.
In contrast, since the Majorana cones are exponentially localized surface modes, we expect them to remain insensitive to changes in $N_z$, provided that the latter remains significantly larger than the localization length of the surface states.
In fact, in our model we find that reducing the thickness down to $N_z=4$ leads to a quasi-2D model in which the only gapless degrees of freedom are the superconducting nodes, as shown in Fig.~\ref{fig:BZ_gap}(c).

As mentioned in the introduction, the Majorana cones are characterized by a winding number \cite{Wen_2002, Beri_2010} taking values of either $+1$ or $-1$.
This topological protection is enabled by the combination of two symmetries, time-reversal symmetry and particle-hole symmetry, placing the system in class DIII according to the Altland-Zirnbauer classification \cite{Altland_1997}.
It is the product of the two symmetries, called chiral symmetry, which enables the topological invariant to be defined.

Both the rotation symmetry of the model as well as time-reversal map the Majorana cones of a given surface onto each other, but these operations do not change the winding number \cite{Beri_2010}.
This makes the surfaces of PtBi$_2$ so-called ``anomalous'' topological superconductors \cite{Changdar_2025}, in the sense that all Majorana cones on the same surface have the same winding number ($+1$ on one surface, and $-1$ on the other).
The superconducting phase of a single surface can therefore not be realized in a purely 2D system, for which the total winding number of all nodes in the BZ should vanish.
In our model, for the chosen global sign of the pairing term, we find that Majorana cones have winding numbers $+1$ on the surface at $z=1$ and $-1$ at $z=N_z$.
The total winding number throughout the slab BZ does vanish, but only when summing the contributions of the Majorana cones on both surfaces.

Taking into account both the symmetry protection of the nodes as well as the distribution of their winding numbers, we identify two mechanisms that will lead to gapping out the Majorana cones.
The first is due to hybridization between cones with opposite winding numbers, which are thus located on opposite surfaces.
This term is exponentially suppressed for sufficiently thick slabs, and becomes relevant only when $N_z$ becomes comparable to the penetration depth of the surface modes.
Further, this term also depends on the relative position of the cone pairs in the slab BZ, leading to a larger gap when the two cones occur at the same momentum $(k_x,k_y)$ and a smaller gap when they occur at different momenta.
To see this, one can imagine the full slab Hamiltonian as being composed of independent, finite 1D systems parameterized by the two variables $k_{x,y}$.
When the same 1D system contains mid-gap states at both ends (same-momentum nodes), their hybridization gap will be enhanced, since it is proportional to the overlap of the end-mode wavefunctions.
This means that hybridization gaps can be controlled not just by slab thickness, but also by varying the surface potential $\mu_\text{surf}$, which has the effect of shifting the position of the Majorana cones in the BZ.

The second mechanism for gapping out the Majorana cones relies on breaking the symmetry responsible for their protection, namely the product between time-reversal and particle-hole symmetry.
We achieve this by adding a Zeeman field to the slab Hamiltonian,
\begin{equation}\label{eq:ham_with_Zeeman}
    H_\text{slab} \to H_\text{slab} + V_z \tau_0 \sigma_z \mathbb{1}_{N_z}.
\end{equation}
Here, $\mathbb{1}_{N_z}$ is an identity matrix of the same size as the number of unit cells in the slab, and the juxtaposition of matrices implies a Kronecker product, as was the case with Pauli matrices $\tau_j$ and $\sigma_j$.
According to Eq.~\eqref{eq:ham_with_Zeeman}, the Zeeman field is perpendicular to the surfaces of the slab, it is present throughout its entire thickness, and its strength is given by the Zeeman energy $V_z$.
Unless otherwise indicated, we will use a value $V_z=0.15$ throughout.

\begin{figure}[tb]
\centering
\includegraphics[width=\columnwidth]{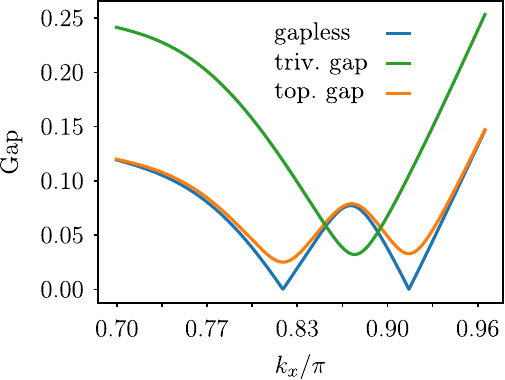}
\caption{
Gap of a thin slab ($N_z = 4$), in units of the intra-orbital hopping $t$, as a function of $k_x$ along the line $k_y = 0$. The blue line shows the gapless Majorana cones in the case of $\mu_\text{surf} = 0.3$ and $V_z = 0$. The orange and green line show the two mechanisms that can lead to a gap appearing : either by removing the chemical potential shift such that $\mu_\text{surf}=V_z=0$ (green line) or by keeping $\mu_{\text{surf}} = 0.3$ and adding a non-zero Zeeman field ($V_z=0.15$, orange line). }
\label{fig:Majorana_gapping}
\end{figure}

In Fig.~\ref{fig:Majorana_gapping} we show that both mechanisms (taking $\mu_\text{surf} = 0$ or adding a Zeeman field $V_z \neq 0$) lead to a gap in the system's spectrum.
The gap obtained by hybridization of opposite Majorana cones is trivial in nature.
In a geometry in which the quasi-2D thin film is finite in both the $x$ and $y$ directions, no topological mid-gap states occur either in the bulk or at the boundary.

\section{Chern superconductor}
\label{sec:chern_sc}

The Zeeman field of Eq.~\eqref{eq:ham_with_Zeeman} leads to a nontrivial gap in the Majorana cones, such that the quasi-2D slab is characterized by a nonzero Chern number.
This is analogous to how the time-reversal symmetry breaking Haldane term gaps out the Dirac cones of the honeycomb lattice \cite{Haldane_1988}, resulting in a Chern insulating phase.
In the case of the toy model Eq.~\eqref{eq:BdG_ham}, when the Zeeman gap is larger than the trivial one, each pair of Majorana cones contributes a value of one to the total Chern number, such that the total invariant of the slab Hamiltonian is either $+6$ or $-6$, depending on the sign of $V_z$.

\begin{figure}[tb]
\centering
\includegraphics[width=\linewidth]{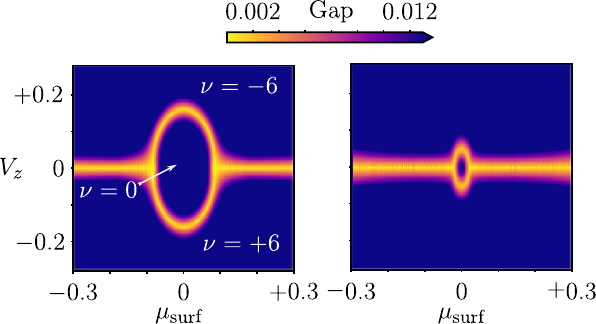}
\caption{
Gap minimum as a function of the Zeeman energy $V_z$ and the surface potential $\mu_{\text{surf}}$, shown for a slab of thickness $N_z=4$ (left) and $N_z=7$ (right). 
The three different regions correspond to a Chern number of $-6$ (above), $0$ (center) and $6$ (below). 
Increasing the slab thickness decreases the parameter region where the trivial phase is found, since the overlap between Majorana cones on opposite surfaces is exponentially suppressed as a function of $N_z$.}
\label{fig:phase_diag}
\end{figure}

The Chern number is defined as \cite{Avron_1983, Bellissard_1994}
\begin{equation}\label{eq:Chern}
\nu=\frac{1}{2\pi i}\int_{\mathrm{BZ}} d^2k\,
\text{Tr}\Big(
P(\mathbf k)
\left[
\partial_{k_x}P(\mathbf k),
\partial_{k_y}P(\mathbf k)
\right]
\Big),
\end{equation}
where Tr denotes the trace and
\begin{equation}
    P(\mathbf k)=\sum_{n \ \text{with} \ E_n<0}
|u_n(\mathbf k)\rangle\langle u_n(\mathbf k)|
\end{equation}
is the projector onto Bloch eigenstates $|u_n(\mathbf k)\rangle$ with energies below zero.
Numerically, we compute the Chern number by discretizing the BZ using the method introduced by Fukui, Hatsugai, and Suzuki \cite{Fukui_2005}, specifically the multi-band formula.
The resulting topological phase diagram of the slab as a function of both $\mu_\text{surf}$ and $V_z$ is shown in Fig.~\ref{fig:phase_diag}, where the left and right panels correspond to thicknesses $N_z=4$ and $7$, respectively.
For numerical efficiency, we have plotted the value of the minimum gap over the slab BZ, as detailed in Appendix \ref{app:numerics}, and then computed the Chern number at one point in each of the gapped phases.

As shown in Fig.~\ref{fig:phase_diag}, for large values of $\mu_\text{surf}$ the Majorana cones remain gapless (bright horizontal lines), due to the fact that they occur at different momenta in the slab BZ.
The Zeeman field then leads to a nontrivial system characterized by Chern numbers $\nu=\pm 6$, depending on its orientation, i.e.~on the sign of $V_z$.
In contrast, for small values of $\mu_\text{surf}$, inter-cone hybridization dominates, producing a trivial superconductor.
Reaching a topological phase in this regime requires a Zeeman field which is strong enough to overcome the trivial gap, leading to the formation of a $\nu=0$ ``bubble'' in the center of the phase diagram.
Note also that increasing the thickness of the slab reduces the overlap of the Majorana cone wavefunctions on the different surfaces, such that the bubble for $N_z=7$ (right panel) is smaller than that for $N_z=4$ (left panel).

\section{Chiral Majorana edge modes}
\label{sec:chiral_edge}

\begin{figure*}[tb]
\centering
\includegraphics[width=0.9\linewidth]{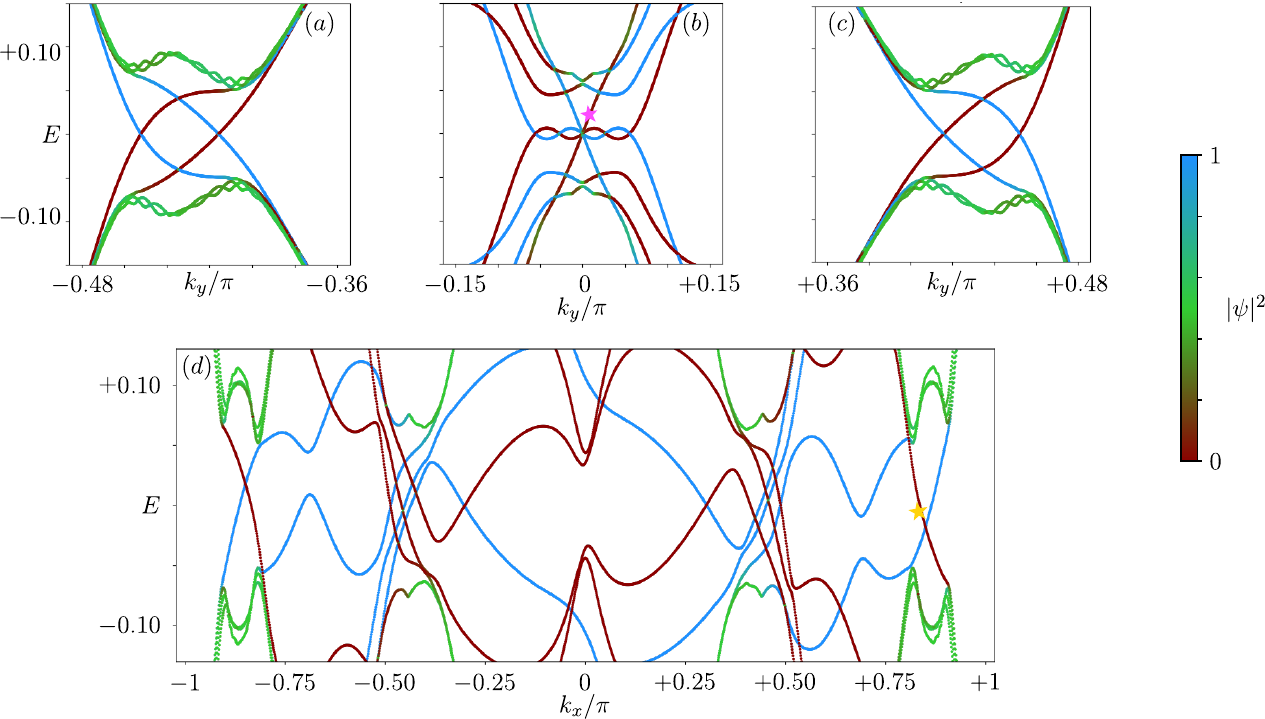}
\caption{
Panels (a,b,c): Parts of the spectrum of a prism infinite along the $y$ direction where Majorana cones appear [compare with the vertical axis in Fig.~\ref{fig:BZ_gap}(a)]. 
Panel (a): Negative $k_y$ part of the prism BZ, corresponding to the bottom part of Fig.~\ref{fig:BZ_gap}.
Panel (b): center part if the prism BZ, corresponding to the middle part of Fig.~\ref{fig:BZ_gap}(a).
Panel (c): positive $k_y$ part of the prism BZ, corresponding to the upper part of Fig.~\ref{fig:BZ_gap}(a).
Panel (d): Spectrum of a prism infinite along the $x$ direction [compare with the horizontal axis in Fig.~\ref{fig:BZ_gap}(a)]. 
The net number of edge states can be seen to be $6$ in either case. The color scale denotes the probability density of a given state, summed over all internal degrees of freedom, all $z$ coordinates, and half of the unit cells in the finite horizontal direction ($x$ for figures in the top row, $y$ in the bottom row).
Therefore, extended 2D states are shown in green, whereas states localized on the different side boundaries of the slab are shown in red and blue, respectively.
In all panels, $N_z = 4$, $M=150$, $V_z=0.15$ and only the 8 eigenvalues closest to $E=0$ have been computed.
}
\label{fig:spectrum}
\end{figure*}

The hallmark of a Chern superconductor is the presence of 1D mid-gap Majorana modes, which propagate unidirectionally along the boundary of the 2D system.
We investigate the properties of these chiral modes by considering a thin prism geometry:
The system is finite both in the $z$ direction ($N_z$ unit cells) and one in plane direction ($M$ unit cells along either $x$ or $y$).
Plotting the bandstructures of prisms infinite in either the $x$ or $y$ direction in Fig.~\ref{fig:spectrum}, we see that indeed the Chern number $\nu=6$ is associated to six chiral Majorana modes.
Here, the color scale denotes the probability density of each state summed over all internal degrees of freedom, over the entire perpendicular ($z$) direction, and over half of the horizontal direction ($x$ or $y$) in which the slab is finite. Due to this choice, bulk modes are plotted in green and chiral mid-gap states on opposite edges are shown in dark red and blue, respectively.

Upon setting $\mu_\text{surf}=0.3$ and gradually increasing $V_z$, the Majorana flat bands occurring on the side-edges of the system in the $V_z=0$ limit \cite{Changdar_2025} acquire nonzero velocities and evolve into chiral Majorana modes.
As such, the unidirectional mid-gap states fully interpolate between the projections of the (now gapped) Majorana cones on the edge BZ, e.g.~as in Fig.~\ref{fig:spectrum}(a) and (c).
When the system is finite in the $x$ direction, such that only $k_y$ remains a good quantum number, pairs of Majorana cones are projected on top of each other, as can be seen also from the cone positions in Fig.~\ref{fig:BZ_gap}(a).
Thus, two chiral modes occur on each edge around $k_y \sim \pm 0.5\pi$, shown in panels (a) and (c).
At $k_y=0$ instead, four Majorana cones are projected onto the same point, two cones with winding number $+1$ and two with $-1$.
We observe that while the net number of chiral edge modes remains the same (panel b), since it is dictated by the Chern number, multiple additional pairs of counter-propagating in-gap edge modes are also formed.
The presence of additional counter-propagating edge states is even more pronounced for the prism geometry which is finite in the $y$ direction (panel d): While there are a net number of $6$ chiral edge modes as required by the Chern number, multiple other counter-propagating pairs can be seen.
Only the gapped Majorana cones around $k_x = \pm 0.9 \pi$ (outer plot regions) host a clean single chiral mode interpolating between the gapped cones.
Note that this is the minimal number due to the fact that outer Majorana cones do not project on top of each other in the edge BZ in this case [compare with Fig.~\ref{fig:BZ_gap}(a)].

Both the chiral modes ($V_z \neq 0$) and the dispersionless Majorana edge states ($V_z=0$) gradually merge with the 2D bulk states of the thin slab as they approach the momenta of the Majorana cones.
This is analogous to how 2D Fermi arcs gradually merge with 3D Weyl states as they approach the momenta of the bulk cones.
We therefore expect that the minimal localization length of the 1D chiral modes depends on the momentum space separation between the Majorana cone projections on the edge BZ, and thus on edge orientation.

\begin{figure}[tb]
\centering
\includegraphics[width=0.9\columnwidth]{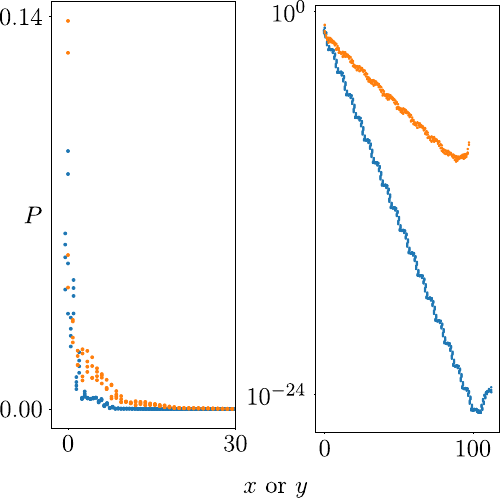}
\caption{
Left panel: Probability distribution of two chiral edge states as a function of position in the direction perpendicular to the edge (in units of the lattice constant). 
The blue color corresponds to the edge state indicated by a yellow star in panel Fig.~\ref{fig:spectrum}(d), and the orange color corresponds to the edge state indicated by a purple star in  Fig.~\ref{fig:spectrum}(b).
Right panel: Same plot but on a semi-log scale.
By fitting with an exponentially decaying function, we obtain localization lengths $\xi_y = 1.92$, and $\xi_x = 5.09$.}
\label{fig:loc_len}
\end{figure}

We examine the real-space profile of edge-state wavefunctions in Fig.~\ref{fig:loc_len}, which shows the probability density of two chiral Majorana modes (blue and orange).
This probability density is obtained by summing over all internal degrees of freedom $\alpha$ (both orbital and spin), as well as summing over all $z$ coordinates.
For example, when the prism is finite in the $x$ direction, an edge mode positioned at momentum $k_{y,0}$ with wavefunction $\psi$ will be associated with a probability density
\begin{equation}
    P(x;k_{y,0}) = \sum_{\alpha, z} | \psi(k_{y,0},x, z, \alpha) |^2.
\end{equation}
The blue color in Fig.~\ref{fig:loc_len} corresponds to the wavefunction of the chiral mode close to $k_x=0.9\pi$ [yellow star in Fig.~\ref{fig:spectrum}(d)], for which the Majorana cones project onto different momenta on the edge BZ.
The orange color corresponds to the edge mode represented by an purple star in Fig.~\ref{fig:spectrum}(b):
Here the Majorana cone projections overlap at $k_y=0$.
Their localization lengths in the direction perpendicular to the edge are clearly different, as can be seen both in the linear as well as in the semilog plot.
By fitting with an exponentially decaying function, $\propto \exp(-x/\xi)$, we find localization lengths that differ by more than a factor of two on the two edges.

\section{Conclusion}
\label{sec:conc}

We have studied the toy-model Hamiltonian Eq.~\eqref{eq:ham_general} to predict some of the features expected to be observed in the candidate topological superconductor PtBi$_2$.
The bulk Weyl cones are gapped out by a finite-size effect (reducing slab thickness), whereas the surface Majorana cones can be gapped out by two competing mechanisms:
Inter-cone hybridization promotes the formation of a trivial gap, whereas time-reversal symmetry breaking (due to a Zeeman field) leads to a nontrivial gap.
In the latter case, six chiral edge modes form on the side-edges of the slab, and their localization length depends on the edge orientation.

The simplicity of the model, while allowing for the study of system sizes normally out of reach for more realistic Wannier Hamiltonians, does introduce a caveat: namely neglecting all other (trivial) bulk states except for the Weyl cones.
Our findings thus require that these extra bulk states do not overlap with and obscure the chiral Majorana edge modes in the edge BZ.
One might however expect that finite size effects due to the slab's thickness should introduce a gap in the bulk states, thus plausibly making the physics described in this work accessible.
Alternatively, one can envisage lowering the temperature below $0.5$ K \cite{Zhang_2025}, causing some or all of the extra bulk states to be gapped out by superconductivity.

We expect that the properties we have examined will be relevant to future PtBi$_2$ experiments due to the fact that they occur only due to nontrivial topology, and as such are independent of other material details.
This includes the gapping out of bulk Weyl cones and surface Majorana cones, which may be probed using angle-resolved photoemission spectroscopy \cite{Kuibarov_2024, Changdar_2025, arxiv.2607.01947, arxiv.2607.26804}, or scanning tunneling spectroscopy \cite{Schimmel_2024, Moreno_2026}.
Further, our prediction that the chiral Majorana edge states should have different thickness on different edges of the quasi-2D system could potentially be observed in scanning tunneling microscopy experiments, as reported in Refs.~\cite{Palacio_Morales_2019, Jiao_2020}.

Beyond controlling the sample thickness, manipulating the topological superconducting phase could be achieved either by means of a Zeeman field or by changing the surface potential in order to move the Majorana cones in momentum space.
Possibilities include applying an external magnetic field, gating the surface, or placing the surface in proximity to a ferromagnet. 
The latter might both induce a Zeeman field as well as cause charge transfer to the surface.
In particular, we expect that changing the surface potential may have a strong effect on inter-cone hybridization in PtBi$_2$, since ab-initio calculations have found that Fermi arcs on opposite surfaces have a momentum-space separation which is much smaller than the size of the BZ \cite{Vocaturo_2024, Changdar_2025, arxiv.2607.01947}.

\section*{Acknowledgments}

This work was supported by the Deutsche Forschungsgemeinschaft (DFG, German Research Foundation) under Germany's Excellence Strategy through the W\"urzburg-Dresden Cluster of Excellence ctd.qmat (EXC 2147, project id 390858490).

\section*{Data Availability}

Our code and the data generated as part of this work are available on Zenodo at \cite{Zenodo_code}.

\appendix

\section{Phase diagram numerical details}
\label{app:numerics}

To produce the phase diagram shown in Fig.~\ref{fig:phase_diag}, we have computed the gap minimum for different values of the parameters $V_z$ and $\mu_\text{surf}$.
Given the computational cost of obtaining the diagram at high resolution, we have not computed the gap over the entire BZ, opting instead to take advantage of the symmetries of the model.
We restricted the gap computation to a line in the 2D BZ: along the $k_y=0$ line, and for $k_x \in [0.80 , 0.93] \pi$.
This is justified as we know, based on the symmetries of the model (e.g.~see Fig.~\ref{fig:Majorana_gapping}),
that this is the region where two Majorana cones appear, i.e.~that the lowest-energy degrees of freedom occur in this interval.
Since the gap opening/closing occurs at the Majorana cones, and since all other nodes are related via symmetries to the two appearing in this momentum interval, we are thus able to obtain complete information on the gap closing in parameter space over the entire 2D BZ.
We note that we have checked the correctness of the resulting phase diagram in a lower-resolution simulation that scans over the full BZ. 

Finally, having obtained a map of the gap closings in parameter space, it sufficed to compute the Chern number once in each gapped region.
To do this, since the slab is a multi-band system, we employed the algorithm described in \cite{Fukui_2005}.

\clearpage

\bibliography{bib}

\begin{thebibliography}{54}%
\makeatletter
\providecommand \@ifxundefined [1]{%
 \@ifx{#1\undefined}
}%
\providecommand \@ifnum [1]{%
 \ifnum #1\expandafter \@firstoftwo
 \else \expandafter \@secondoftwo
 \fi
}%
\providecommand \@ifx [1]{%
 \ifx #1\expandafter \@firstoftwo
 \else \expandafter \@secondoftwo
 \fi
}%
\providecommand \natexlab [1]{#1}%
\providecommand \enquote  [1]{``#1''}%
\providecommand \bibnamefont  [1]{#1}%
\providecommand \bibfnamefont [1]{#1}%
\providecommand \citenamefont [1]{#1}%
\providecommand \href@noop [0]{\@secondoftwo}%
\providecommand \href [0]{\begingroup \@sanitize@url \@href}%
\providecommand \@href[1]{\@@startlink{#1}\@@href}%
\providecommand \@@href[1]{\endgroup#1\@@endlink}%
\providecommand \@sanitize@url [0]{\catcode `\\12\catcode `\$12\catcode
  `\&12\catcode `\#12\catcode `\^12\catcode `\_12\catcode `\%12\relax}%
\providecommand \@@startlink[1]{}%
\providecommand \@@endlink[0]{}%
\providecommand \url  [0]{\begingroup\@sanitize@url \@url }%
\providecommand \@url [1]{\endgroup\@href {#1}{\urlprefix }}%
\providecommand \urlprefix  [0]{URL }%
\providecommand \Eprint [0]{\href }%
\providecommand \doibase [0]{https://doi.org/}%
\providecommand \selectlanguage [0]{\@gobble}%
\providecommand \bibinfo  [0]{\@secondoftwo}%
\providecommand \bibfield  [0]{\@secondoftwo}%
\providecommand \translation [1]{[#1]}%
\providecommand \BibitemOpen [0]{}%
\providecommand \bibitemStop [0]{}%
\providecommand \bibitemNoStop [0]{.\EOS\space}%
\providecommand \EOS [0]{\spacefactor3000\relax}%
\providecommand \BibitemShut  [1]{\csname bibitem#1\endcsname}%
\let\auto@bib@innerbib\@empty
\bibitem [{\citenamefont {Alicea}(2012)}]{Alicea_2012}%
  \BibitemOpen
  \bibfield  {author} {\bibinfo {author} {\bibfnamefont {J.}~\bibnamefont
  {Alicea}},\ }\bibfield  {title} {\bibinfo {title} {{New directions in the
  pursuit of Majorana fermions in solid state systems}},\ }\href
  {https://doi.org/10.1088/0034-4885/75/7/076501} {\bibfield  {journal}
  {\bibinfo  {journal} {Rep. Prog. Phys.}\ }\textbf {\bibinfo {volume} {75}},\
  \bibinfo {pages} {076501} (\bibinfo {year} {2012})}\BibitemShut {NoStop}%
\bibitem [{\citenamefont {Beenakker}(2013)}]{Beenakker_2013}%
  \BibitemOpen
  \bibfield  {author} {\bibinfo {author} {\bibfnamefont {C.}~\bibnamefont
  {Beenakker}},\ }\bibfield  {title} {\bibinfo {title} {{Search for Majorana
  Fermions in Superconductors}},\ }\href
  {https://doi.org/10.1146/annurev-conmatphys-030212-184337} {\bibfield
  {journal} {\bibinfo  {journal} {Annu. Rev. Condens. Matter Phys.}\ }\textbf
  {\bibinfo {volume} {4}},\ \bibinfo {pages} {113} (\bibinfo {year}
  {2013})}\BibitemShut {NoStop}%
\bibitem [{\citenamefont {Sato}\ and\ \citenamefont
  {Fujimoto}(2016)}]{Sato_2016}%
  \BibitemOpen
  \bibfield  {author} {\bibinfo {author} {\bibfnamefont {M.}~\bibnamefont
  {Sato}}\ and\ \bibinfo {author} {\bibfnamefont {S.}~\bibnamefont
  {Fujimoto}},\ }\bibfield  {title} {\bibinfo {title} {{Majorana Fermions and
  Topology in Superconductors}},\ }\href
  {https://doi.org/10.7566/jpsj.85.072001} {\bibfield  {journal} {\bibinfo
  {journal} {J. Phys. Soc. Jpn.}\ }\textbf {\bibinfo {volume} {85}},\ \bibinfo
  {pages} {072001} (\bibinfo {year} {2016})}\BibitemShut {NoStop}%
\bibitem [{\citenamefont {Sato}\ and\ \citenamefont {Ando}(2017)}]{Sato_2017}%
  \BibitemOpen
  \bibfield  {author} {\bibinfo {author} {\bibfnamefont {M.}~\bibnamefont
  {Sato}}\ and\ \bibinfo {author} {\bibfnamefont {Y.}~\bibnamefont {Ando}},\
  }\bibfield  {title} {\bibinfo {title} {{Topological superconductors: a
  review}},\ }\href {https://doi.org/10.1088/1361-6633/aa6ac7} {\bibfield
  {journal} {\bibinfo  {journal} {Rep. Prog. Phys.}\ }\textbf {\bibinfo
  {volume} {80}},\ \bibinfo {pages} {076501} (\bibinfo {year}
  {2017})}\BibitemShut {NoStop}%
\bibitem [{\citenamefont {Sharma}\ \emph {et~al.}(2022)\citenamefont {Sharma},
  \citenamefont {Sharma}, \citenamefont {Karn},\ and\ \citenamefont
  {Awana}}]{Sharma_2022}%
  \BibitemOpen
  \bibfield  {author} {\bibinfo {author} {\bibfnamefont {M.~M.}\ \bibnamefont
  {Sharma}}, \bibinfo {author} {\bibfnamefont {P.}~\bibnamefont {Sharma}},
  \bibinfo {author} {\bibfnamefont {N.~K.}\ \bibnamefont {Karn}},\ and\
  \bibinfo {author} {\bibfnamefont {V.~P.~S.}\ \bibnamefont {Awana}},\
  }\bibfield  {title} {\bibinfo {title} {{Comprehensive review on topological
  superconducting materials and interfaces}},\ }\href
  {https://doi.org/10.1088/1361-6668/ac6987} {\bibfield  {journal} {\bibinfo
  {journal} {Supercond. Sci. Technol.}\ }\textbf {\bibinfo {volume} {35}},\
  \bibinfo {pages} {083003} (\bibinfo {year} {2022})}\BibitemShut {NoStop}%
\bibitem [{\citenamefont {Kitaev}(2001)}]{Kitaev_2001}%
  \BibitemOpen
  \bibfield  {author} {\bibinfo {author} {\bibfnamefont {A.~Y.}\ \bibnamefont
  {Kitaev}},\ }\bibfield  {title} {\bibinfo {title} {{Unpaired Majorana
  fermions in quantum wires}},\ }\href
  {https://doi.org/10.1070/1063-7869/44/10s/s29} {\bibfield  {journal}
  {\bibinfo  {journal} {Phys.-Usp.}\ }\textbf {\bibinfo {volume} {44}},\
  \bibinfo {pages} {131} (\bibinfo {year} {2001})}\BibitemShut {NoStop}%
\bibitem [{\citenamefont {Schnyder}\ \emph {et~al.}(2008)\citenamefont
  {Schnyder}, \citenamefont {Ryu}, \citenamefont {Furusaki},\ and\
  \citenamefont {Ludwig}}]{Schnyder_2008}%
  \BibitemOpen
  \bibfield  {author} {\bibinfo {author} {\bibfnamefont {A.~P.}\ \bibnamefont
  {Schnyder}}, \bibinfo {author} {\bibfnamefont {S.}~\bibnamefont {Ryu}},
  \bibinfo {author} {\bibfnamefont {A.}~\bibnamefont {Furusaki}},\ and\
  \bibinfo {author} {\bibfnamefont {A.~W.~W.}\ \bibnamefont {Ludwig}},\
  }\bibfield  {title} {\bibinfo {title} {{Classification of topological
  insulators and superconductors in three spatial dimensions}},\ }\href
  {https://doi.org/10.1103/physrevb.78.195125} {\bibfield  {journal} {\bibinfo
  {journal} {Phys. Rev. B}\ }\textbf {\bibinfo {volume} {78}},\ \bibinfo
  {pages} {195125} (\bibinfo {year} {2008})}\BibitemShut {NoStop}%
\bibitem [{\citenamefont {Qi}\ \emph {et~al.}(2010)\citenamefont {Qi},
  \citenamefont {Hughes},\ and\ \citenamefont {Zhang}}]{Qi_2010}%
  \BibitemOpen
  \bibfield  {author} {\bibinfo {author} {\bibfnamefont {X.-L.}\ \bibnamefont
  {Qi}}, \bibinfo {author} {\bibfnamefont {T.~L.}\ \bibnamefont {Hughes}},\
  and\ \bibinfo {author} {\bibfnamefont {S.-C.}\ \bibnamefont {Zhang}},\
  }\bibfield  {title} {\bibinfo {title} {{Topological invariants for the Fermi
  surface of a time-reversal-invariant superconductor}},\ }\href
  {https://doi.org/10.1103/physrevb.81.134508} {\bibfield  {journal} {\bibinfo
  {journal} {Phys. Rev. B}\ }\textbf {\bibinfo {volume} {81}},\ \bibinfo
  {pages} {134508} (\bibinfo {year} {2010})}\BibitemShut {NoStop}%
\bibitem [{\citenamefont {Qi}\ and\ \citenamefont {Zhang}(2011)}]{Qi_2011}%
  \BibitemOpen
  \bibfield  {author} {\bibinfo {author} {\bibfnamefont {X.-L.}\ \bibnamefont
  {Qi}}\ and\ \bibinfo {author} {\bibfnamefont {S.-C.}\ \bibnamefont {Zhang}},\
  }\bibfield  {title} {\bibinfo {title} {{Topological insulators and
  superconductors}},\ }\href {https://doi.org/10.1103/revmodphys.83.1057}
  {\bibfield  {journal} {\bibinfo  {journal} {Rev. Mod. Phys.}\ }\textbf
  {\bibinfo {volume} {83}},\ \bibinfo {pages} {1057} (\bibinfo {year}
  {2011})}\BibitemShut {NoStop}%
\bibitem [{\citenamefont {Sato}\ \emph {et~al.}(2010)\citenamefont {Sato},
  \citenamefont {Takahashi},\ and\ \citenamefont {Fujimoto}}]{Sato_2010}%
  \BibitemOpen
  \bibfield  {author} {\bibinfo {author} {\bibfnamefont {M.}~\bibnamefont
  {Sato}}, \bibinfo {author} {\bibfnamefont {Y.}~\bibnamefont {Takahashi}},\
  and\ \bibinfo {author} {\bibfnamefont {S.}~\bibnamefont {Fujimoto}},\
  }\bibfield  {title} {\bibinfo {title} {{Non-Abelian topological orders and
  Majorana fermions in spin-singlet superconductors}},\ }\href
  {https://doi.org/10.1103/physrevb.82.134521} {\bibfield  {journal} {\bibinfo
  {journal} {Phys. Rev. B}\ }\textbf {\bibinfo {volume} {82}},\ \bibinfo
  {pages} {134521} (\bibinfo {year} {2010})}\BibitemShut {NoStop}%
\bibitem [{\citenamefont {Sau}\ \emph {et~al.}(2010)\citenamefont {Sau},
  \citenamefont {Lutchyn}, \citenamefont {Tewari},\ and\ \citenamefont
  {Das~Sarma}}]{Sau_2010}%
  \BibitemOpen
  \bibfield  {author} {\bibinfo {author} {\bibfnamefont {J.~D.}\ \bibnamefont
  {Sau}}, \bibinfo {author} {\bibfnamefont {R.~M.}\ \bibnamefont {Lutchyn}},
  \bibinfo {author} {\bibfnamefont {S.}~\bibnamefont {Tewari}},\ and\ \bibinfo
  {author} {\bibfnamefont {S.}~\bibnamefont {Das~Sarma}},\ }\bibfield  {title}
  {\bibinfo {title} {{Generic New Platform for Topological Quantum Computation
  Using Semiconductor Heterostructures}},\ }\href
  {https://doi.org/10.1103/physrevlett.104.040502} {\bibfield  {journal}
  {\bibinfo  {journal} {Phys. Rev. Lett.}\ }\textbf {\bibinfo {volume} {104}},\
  \bibinfo {pages} {040502} (\bibinfo {year} {2010})}\BibitemShut {NoStop}%
\bibitem [{\citenamefont {Lutchyn}\ \emph {et~al.}(2010)\citenamefont
  {Lutchyn}, \citenamefont {Sau},\ and\ \citenamefont
  {Das~Sarma}}]{Lutchyn_2010}%
  \BibitemOpen
  \bibfield  {author} {\bibinfo {author} {\bibfnamefont {R.~M.}\ \bibnamefont
  {Lutchyn}}, \bibinfo {author} {\bibfnamefont {J.~D.}\ \bibnamefont {Sau}},\
  and\ \bibinfo {author} {\bibfnamefont {S.}~\bibnamefont {Das~Sarma}},\
  }\bibfield  {title} {\bibinfo {title} {{Majorana Fermions and a Topological
  Phase Transition in Semiconductor-Superconductor Heterostructures}},\ }\href
  {https://doi.org/10.1103/physrevlett.105.077001} {\bibfield  {journal}
  {\bibinfo  {journal} {Phys. Rev. Lett.}\ }\textbf {\bibinfo {volume} {105}},\
  \bibinfo {pages} {077001} (\bibinfo {year} {2010})}\BibitemShut {NoStop}%
\bibitem [{\citenamefont {Mourik}\ \emph {et~al.}(2012)\citenamefont {Mourik},
  \citenamefont {Zuo}, \citenamefont {Frolov}, \citenamefont {Plissard},
  \citenamefont {Bakkers},\ and\ \citenamefont {Kouwenhoven}}]{Mourik_2012}%
  \BibitemOpen
  \bibfield  {author} {\bibinfo {author} {\bibfnamefont {V.}~\bibnamefont
  {Mourik}}, \bibinfo {author} {\bibfnamefont {K.}~\bibnamefont {Zuo}},
  \bibinfo {author} {\bibfnamefont {S.~M.}\ \bibnamefont {Frolov}}, \bibinfo
  {author} {\bibfnamefont {S.~R.}\ \bibnamefont {Plissard}}, \bibinfo {author}
  {\bibfnamefont {E.~P. A.~M.}\ \bibnamefont {Bakkers}},\ and\ \bibinfo
  {author} {\bibfnamefont {L.~P.}\ \bibnamefont {Kouwenhoven}},\ }\bibfield
  {title} {\bibinfo {title} {{Signatures of Majorana Fermions in Hybrid
  Superconductor-Semiconductor Nanowire Devices}},\ }\href
  {https://doi.org/10.1126/science.1222360} {\bibfield  {journal} {\bibinfo
  {journal} {Science}\ }\textbf {\bibinfo {volume} {336}},\ \bibinfo {pages}
  {1003} (\bibinfo {year} {2012})}\BibitemShut {NoStop}%
\bibitem [{\citenamefont {Won}\ \emph {et~al.}(2004)\citenamefont {Won},
  \citenamefont {Parker}, \citenamefont {Haas},\ and\ \citenamefont
  {Maki}}]{Won_2004}%
  \BibitemOpen
  \bibfield  {author} {\bibinfo {author} {\bibfnamefont {H.}~\bibnamefont
  {Won}}, \bibinfo {author} {\bibfnamefont {D.}~\bibnamefont {Parker}},
  \bibinfo {author} {\bibfnamefont {S.}~\bibnamefont {Haas}},\ and\ \bibinfo
  {author} {\bibfnamefont {K.}~\bibnamefont {Maki}},\ }\bibfield  {title}
  {\bibinfo {title} {{Aspects of nodal superconductivity}},\ }\href
  {https://doi.org/10.1016/j.cap.2004.01.011} {\bibfield  {journal} {\bibinfo
  {journal} {Curr. Appl Phys.}\ }\textbf {\bibinfo {volume} {4}},\ \bibinfo
  {pages} {523} (\bibinfo {year} {2004})}\BibitemShut {NoStop}%
\bibitem [{\citenamefont {Wen}\ and\ \citenamefont {Zee}(2002)}]{Wen_2002}%
  \BibitemOpen
  \bibfield  {author} {\bibinfo {author} {\bibfnamefont {X.~G.}\ \bibnamefont
  {Wen}}\ and\ \bibinfo {author} {\bibfnamefont {A.}~\bibnamefont {Zee}},\
  }\bibfield  {title} {\bibinfo {title} {{Gapless fermions and quantum
  order}},\ }\href {https://doi.org/10.1103/physrevb.66.235110} {\bibfield
  {journal} {\bibinfo  {journal} {Phys. Rev. B}\ }\textbf {\bibinfo {volume}
  {66}},\ \bibinfo {pages} {235110} (\bibinfo {year} {2002})}\BibitemShut
  {NoStop}%
\bibitem [{\citenamefont {B\'eri}(2010)}]{Beri_2010}%
  \BibitemOpen
  \bibfield  {author} {\bibinfo {author} {\bibfnamefont {B.}~\bibnamefont
  {B\'eri}},\ }\bibfield  {title} {\bibinfo {title} {{Topologically stable
  gapless phases of time-reversal-invariant superconductors}},\ }\href
  {https://doi.org/10.1103/physrevb.81.134515} {\bibfield  {journal} {\bibinfo
  {journal} {Phys. Rev. B}\ }\textbf {\bibinfo {volume} {81}},\ \bibinfo
  {pages} {134515} (\bibinfo {year} {2010})}\BibitemShut {NoStop}%
\bibitem [{\citenamefont {Sato}\ \emph {et~al.}(2011)\citenamefont {Sato},
  \citenamefont {Tanaka}, \citenamefont {Yada},\ and\ \citenamefont
  {Yokoyama}}]{Sato_2011}%
  \BibitemOpen
  \bibfield  {author} {\bibinfo {author} {\bibfnamefont {M.}~\bibnamefont
  {Sato}}, \bibinfo {author} {\bibfnamefont {Y.}~\bibnamefont {Tanaka}},
  \bibinfo {author} {\bibfnamefont {K.}~\bibnamefont {Yada}},\ and\ \bibinfo
  {author} {\bibfnamefont {T.}~\bibnamefont {Yokoyama}},\ }\bibfield  {title}
  {\bibinfo {title} {{Topology of Andreev bound states with flat dispersion}},\
  }\href {https://doi.org/10.1103/physrevb.83.224511} {\bibfield  {journal}
  {\bibinfo  {journal} {Phys. Rev. B}\ }\textbf {\bibinfo {volume} {83}},\
  \bibinfo {pages} {224511} (\bibinfo {year} {2011})}\BibitemShut {NoStop}%
\bibitem [{\citenamefont {Schnyder}\ and\ \citenamefont
  {Brydon}(2015)}]{Schnyder_2015}%
  \BibitemOpen
  \bibfield  {author} {\bibinfo {author} {\bibfnamefont {A.~P.}\ \bibnamefont
  {Schnyder}}\ and\ \bibinfo {author} {\bibfnamefont {P.~M.~R.}\ \bibnamefont
  {Brydon}},\ }\bibfield  {title} {\bibinfo {title} {{Topological surface
  states in nodal superconductors}},\ }\href
  {https://doi.org/10.1088/0953-8984/27/24/243201} {\bibfield  {journal}
  {\bibinfo  {journal} {J. Phys.: Condens. Matter}\ }\textbf {\bibinfo {volume}
  {27}},\ \bibinfo {pages} {243201} (\bibinfo {year} {2015})}\BibitemShut
  {NoStop}%
\bibitem [{\citenamefont {Nakada}\ \emph {et~al.}(1996)\citenamefont {Nakada},
  \citenamefont {Fujita}, \citenamefont {Dresselhaus},\ and\ \citenamefont
  {Dresselhaus}}]{Nakada_1996}%
  \BibitemOpen
  \bibfield  {author} {\bibinfo {author} {\bibfnamefont {K.}~\bibnamefont
  {Nakada}}, \bibinfo {author} {\bibfnamefont {M.}~\bibnamefont {Fujita}},
  \bibinfo {author} {\bibfnamefont {G.}~\bibnamefont {Dresselhaus}},\ and\
  \bibinfo {author} {\bibfnamefont {M.~S.}\ \bibnamefont {Dresselhaus}},\
  }\bibfield  {title} {\bibinfo {title} {{Edge state in graphene ribbons:
  Nanometer size effect and edge shape dependence}},\ }\href
  {https://doi.org/10.1103/physrevb.54.17954} {\bibfield  {journal} {\bibinfo
  {journal} {Phys. Rev. B}\ }\textbf {\bibinfo {volume} {54}},\ \bibinfo
  {pages} {17954} (\bibinfo {year} {1996})}\BibitemShut {NoStop}%
\bibitem [{\citenamefont {Ryu}\ and\ \citenamefont
  {Hatsugai}(2002)}]{Ryu_2002}%
  \BibitemOpen
  \bibfield  {author} {\bibinfo {author} {\bibfnamefont {S.}~\bibnamefont
  {Ryu}}\ and\ \bibinfo {author} {\bibfnamefont {Y.}~\bibnamefont {Hatsugai}},\
  }\bibfield  {title} {\bibinfo {title} {{Topological Origin of Zero-Energy
  Edge States in Particle-Hole Symmetric Systems}},\ }\href
  {https://doi.org/10.1103/physrevlett.89.077002} {\bibfield  {journal}
  {\bibinfo  {journal} {Phys. Rev. Lett.}\ }\textbf {\bibinfo {volume} {89}},\
  \bibinfo {pages} {077002} (\bibinfo {year} {2002})}\BibitemShut {NoStop}%
\bibitem [{\citenamefont {Delplace}\ \emph {et~al.}(2011)\citenamefont
  {Delplace}, \citenamefont {Ullmo},\ and\ \citenamefont
  {Montambaux}}]{Delplace_2011}%
  \BibitemOpen
  \bibfield  {author} {\bibinfo {author} {\bibfnamefont {P.}~\bibnamefont
  {Delplace}}, \bibinfo {author} {\bibfnamefont {D.}~\bibnamefont {Ullmo}},\
  and\ \bibinfo {author} {\bibfnamefont {G.}~\bibnamefont {Montambaux}},\
  }\bibfield  {title} {\bibinfo {title} {{Zak phase and the existence of edge
  states in graphene}},\ }\href {https://doi.org/10.1103/physrevb.84.195452}
  {\bibfield  {journal} {\bibinfo  {journal} {Phys. Rev. B}\ }\textbf {\bibinfo
  {volume} {84}},\ \bibinfo {pages} {195452} (\bibinfo {year}
  {2011})}\BibitemShut {NoStop}%
\bibitem [{\citenamefont {Veyrat}\ \emph {et~al.}(2023)\citenamefont {Veyrat},
  \citenamefont {Labracherie}, \citenamefont {Bashlakov}, \citenamefont
  {Caglieris}, \citenamefont {Facio}, \citenamefont {Shipunov}, \citenamefont
  {Charvin}, \citenamefont {Acharya}, \citenamefont {Naidyuk}, \citenamefont
  {Giraud}, \citenamefont {van~den Brink}, \citenamefont {B{\"u}chner},
  \citenamefont {Hess}, \citenamefont {Aswartham},\ and\ \citenamefont
  {Dufouleur}}]{Veyrat_2023}%
  \BibitemOpen
  \bibfield  {author} {\bibinfo {author} {\bibfnamefont {A.}~\bibnamefont
  {Veyrat}}, \bibinfo {author} {\bibfnamefont {V.}~\bibnamefont {Labracherie}},
  \bibinfo {author} {\bibfnamefont {D.~L.}\ \bibnamefont {Bashlakov}}, \bibinfo
  {author} {\bibfnamefont {F.}~\bibnamefont {Caglieris}}, \bibinfo {author}
  {\bibfnamefont {J.~I.}\ \bibnamefont {Facio}}, \bibinfo {author}
  {\bibfnamefont {G.}~\bibnamefont {Shipunov}}, \bibinfo {author}
  {\bibfnamefont {T.}~\bibnamefont {Charvin}}, \bibinfo {author} {\bibfnamefont
  {R.}~\bibnamefont {Acharya}}, \bibinfo {author} {\bibfnamefont
  {Y.}~\bibnamefont {Naidyuk}}, \bibinfo {author} {\bibfnamefont
  {R.}~\bibnamefont {Giraud}}, \bibinfo {author} {\bibfnamefont
  {J.}~\bibnamefont {van~den Brink}}, \bibinfo {author} {\bibfnamefont
  {B.}~\bibnamefont {B{\"u}chner}}, \bibinfo {author} {\bibfnamefont
  {C.}~\bibnamefont {Hess}}, \bibinfo {author} {\bibfnamefont {S.}~\bibnamefont
  {Aswartham}},\ and\ \bibinfo {author} {\bibfnamefont {J.}~\bibnamefont
  {Dufouleur}},\ }\bibfield  {title} {\bibinfo {title}
  {{Berezinskii-Kosterlitz-Thouless Transition in the Type-I Weyl Semimetal
  PtBi$_2$}},\ }\href {https://doi.org/10.1021/acs.nanolett.2c04297} {\bibfield
   {journal} {\bibinfo  {journal} {Nano Lett.}\ }\textbf {\bibinfo {volume}
  {23}},\ \bibinfo {pages} {1229} (\bibinfo {year} {2023})}\BibitemShut
  {NoStop}%
\bibitem [{\citenamefont {Kuibarov}\ \emph {et~al.}(2024)\citenamefont
  {Kuibarov}, \citenamefont {Suvorov}, \citenamefont {Vocaturo}, \citenamefont
  {Fedorov}, \citenamefont {Lou}, \citenamefont {Merkwitz}, \citenamefont
  {Voroshnin}, \citenamefont {Facio}, \citenamefont {Koepernik}, \citenamefont
  {Yaresko}, \citenamefont {Shipunov}, \citenamefont {Aswartham}, \citenamefont
  {van~den Brink}, \citenamefont {B{\"u}chner},\ and\ \citenamefont
  {Borisenko}}]{Kuibarov_2024}%
  \BibitemOpen
  \bibfield  {author} {\bibinfo {author} {\bibfnamefont {A.}~\bibnamefont
  {Kuibarov}}, \bibinfo {author} {\bibfnamefont {O.}~\bibnamefont {Suvorov}},
  \bibinfo {author} {\bibfnamefont {R.}~\bibnamefont {Vocaturo}}, \bibinfo
  {author} {\bibfnamefont {A.}~\bibnamefont {Fedorov}}, \bibinfo {author}
  {\bibfnamefont {R.}~\bibnamefont {Lou}}, \bibinfo {author} {\bibfnamefont
  {L.}~\bibnamefont {Merkwitz}}, \bibinfo {author} {\bibfnamefont
  {V.}~\bibnamefont {Voroshnin}}, \bibinfo {author} {\bibfnamefont {J.~I.}\
  \bibnamefont {Facio}}, \bibinfo {author} {\bibfnamefont {K.}~\bibnamefont
  {Koepernik}}, \bibinfo {author} {\bibfnamefont {A.}~\bibnamefont {Yaresko}},
  \bibinfo {author} {\bibfnamefont {G.}~\bibnamefont {Shipunov}}, \bibinfo
  {author} {\bibfnamefont {S.}~\bibnamefont {Aswartham}}, \bibinfo {author}
  {\bibfnamefont {J.}~\bibnamefont {van~den Brink}}, \bibinfo {author}
  {\bibfnamefont {B.}~\bibnamefont {B{\"u}chner}},\ and\ \bibinfo {author}
  {\bibfnamefont {S.}~\bibnamefont {Borisenko}},\ }\bibfield  {title} {\bibinfo
  {title} {{Evidence of superconducting Fermi arcs}},\ }\href
  {https://doi.org/10.1038/s41586-023-06977-7} {\bibfield  {journal} {\bibinfo
  {journal} {Nature}\ }\textbf {\bibinfo {volume} {626}},\ \bibinfo {pages}
  {294} (\bibinfo {year} {2024})}\BibitemShut {NoStop}%
\bibitem [{\citenamefont {Schimmel}\ \emph {et~al.}(2024)\citenamefont
  {Schimmel}, \citenamefont {Fasano}, \citenamefont {Hoffmann}, \citenamefont
  {Besproswanny}, \citenamefont {Corredor~Bohorquez}, \citenamefont {Puig},
  \citenamefont {Elshalem}, \citenamefont {Kalisky}, \citenamefont {Shipunov},
  \citenamefont {Baumann}, \citenamefont {Aswartham}, \citenamefont
  {B{\"u}chner},\ and\ \citenamefont {Hess}}]{Schimmel_2024}%
  \BibitemOpen
  \bibfield  {author} {\bibinfo {author} {\bibfnamefont {S.}~\bibnamefont
  {Schimmel}}, \bibinfo {author} {\bibfnamefont {Y.}~\bibnamefont {Fasano}},
  \bibinfo {author} {\bibfnamefont {S.}~\bibnamefont {Hoffmann}}, \bibinfo
  {author} {\bibfnamefont {J.}~\bibnamefont {Besproswanny}}, \bibinfo {author}
  {\bibfnamefont {L.~T.}\ \bibnamefont {Corredor~Bohorquez}}, \bibinfo {author}
  {\bibfnamefont {J.}~\bibnamefont {Puig}}, \bibinfo {author} {\bibfnamefont
  {B.-C.}\ \bibnamefont {Elshalem}}, \bibinfo {author} {\bibfnamefont
  {B.}~\bibnamefont {Kalisky}}, \bibinfo {author} {\bibfnamefont
  {G.}~\bibnamefont {Shipunov}}, \bibinfo {author} {\bibfnamefont
  {D.}~\bibnamefont {Baumann}}, \bibinfo {author} {\bibfnamefont
  {S.}~\bibnamefont {Aswartham}}, \bibinfo {author} {\bibfnamefont
  {B.}~\bibnamefont {B{\"u}chner}},\ and\ \bibinfo {author} {\bibfnamefont
  {C.}~\bibnamefont {Hess}},\ }\bibfield  {title} {\bibinfo {title} {{Surface
  superconductivity in the topological Weyl semimetal t-PtBi$_2$}},\ }\href
  {https://doi.org/10.1038/s41467-024-54389-6} {\bibfield  {journal} {\bibinfo
  {journal} {Nat. Commun.}\ }\textbf {\bibinfo {volume} {15}},\ \bibinfo
  {pages} {9895} (\bibinfo {year} {2024})}\BibitemShut {NoStop}%
\bibitem [{\citenamefont {Changdar}\ \emph {et~al.}(2025)\citenamefont
  {Changdar}, \citenamefont {Suvorov}, \citenamefont {Kuibarov}, \citenamefont
  {Thirupathaiah}, \citenamefont {Shipunov}, \citenamefont {Aswartham},
  \citenamefont {Wurmehl}, \citenamefont {Kovalchuk}, \citenamefont
  {Koepernik}, \citenamefont {Timm}, \citenamefont {B{\"u}chner}, \citenamefont
  {Fulga}, \citenamefont {Borisenko},\ and\ \citenamefont {van~den
  Brink}}]{Changdar_2025}%
  \BibitemOpen
  \bibfield  {author} {\bibinfo {author} {\bibfnamefont {S.}~\bibnamefont
  {Changdar}}, \bibinfo {author} {\bibfnamefont {O.}~\bibnamefont {Suvorov}},
  \bibinfo {author} {\bibfnamefont {A.}~\bibnamefont {Kuibarov}}, \bibinfo
  {author} {\bibfnamefont {S.}~\bibnamefont {Thirupathaiah}}, \bibinfo {author}
  {\bibfnamefont {G.}~\bibnamefont {Shipunov}}, \bibinfo {author}
  {\bibfnamefont {S.}~\bibnamefont {Aswartham}}, \bibinfo {author}
  {\bibfnamefont {S.}~\bibnamefont {Wurmehl}}, \bibinfo {author} {\bibfnamefont
  {I.}~\bibnamefont {Kovalchuk}}, \bibinfo {author} {\bibfnamefont
  {K.}~\bibnamefont {Koepernik}}, \bibinfo {author} {\bibfnamefont
  {C.}~\bibnamefont {Timm}}, \bibinfo {author} {\bibfnamefont {B.}~\bibnamefont
  {B{\"u}chner}}, \bibinfo {author} {\bibfnamefont {I.~C.}\ \bibnamefont
  {Fulga}}, \bibinfo {author} {\bibfnamefont {S.}~\bibnamefont {Borisenko}},\
  and\ \bibinfo {author} {\bibfnamefont {J.}~\bibnamefont {van~den Brink}},\
  }\bibfield  {title} {\bibinfo {title} {{Topological nodal i-wave
  superconductivity in PtBi$_2$}},\ }\href
  {https://doi.org/10.1038/s41586-025-09712-6} {\bibfield  {journal} {\bibinfo
  {journal} {Nature}\ }\textbf {\bibinfo {volume} {647}},\ \bibinfo {pages}
  {613} (\bibinfo {year} {2025})}\BibitemShut {NoStop}%
\bibitem [{\citenamefont {Moreno}\ \emph {et~al.}(2026)\citenamefont {Moreno},
  \citenamefont {Talavera}, \citenamefont {Herrera}, \citenamefont {Valle},
  \citenamefont {Li}, \citenamefont {Wang}, \citenamefont {Bud'ko},
  \citenamefont {Buzdin}, \citenamefont {Guillam{\'o}n}, \citenamefont
  {Canfield},\ and\ \citenamefont {Suderow}}]{Moreno_2026}%
  \BibitemOpen
  \bibfield  {author} {\bibinfo {author} {\bibfnamefont {J.~A.}\ \bibnamefont
  {Moreno}}, \bibinfo {author} {\bibfnamefont {P.~G.}\ \bibnamefont
  {Talavera}}, \bibinfo {author} {\bibfnamefont {E.}~\bibnamefont {Herrera}},
  \bibinfo {author} {\bibfnamefont {S.~L.}\ \bibnamefont {Valle}}, \bibinfo
  {author} {\bibfnamefont {Z.}~\bibnamefont {Li}}, \bibinfo {author}
  {\bibfnamefont {L.-L.}\ \bibnamefont {Wang}}, \bibinfo {author}
  {\bibfnamefont {S.}~\bibnamefont {Bud'ko}}, \bibinfo {author} {\bibfnamefont
  {A.~I.}\ \bibnamefont {Buzdin}}, \bibinfo {author} {\bibfnamefont
  {I.}~\bibnamefont {Guillam{\'o}n}}, \bibinfo {author} {\bibfnamefont {P.~C.}\
  \bibnamefont {Canfield}},\ and\ \bibinfo {author} {\bibfnamefont
  {H.}~\bibnamefont {Suderow}},\ }\bibfield  {title} {\bibinfo {title} {{Robust
  Two-Dimensional Surface Superconductivity and Vortex Lattice in the Weyl
  Semimetal $\gamma$ - PtBi$_2$}},\ }\href {https://doi.org/10.1103/9cyw-m5zr}
  {\bibfield  {journal} {\bibinfo  {journal} {Phys. Rev. Lett.}\ }\textbf
  {\bibinfo {volume} {137}},\ \bibinfo {pages} {086001} (\bibinfo {year}
  {2026})}\BibitemShut {NoStop}%
\bibitem [{\citenamefont {Burkov}\ \emph {et~al.}(2011)\citenamefont {Burkov},
  \citenamefont {Hook},\ and\ \citenamefont {Balents}}]{Burkov_2011}%
  \BibitemOpen
  \bibfield  {author} {\bibinfo {author} {\bibfnamefont {A.~A.}\ \bibnamefont
  {Burkov}}, \bibinfo {author} {\bibfnamefont {M.~D.}\ \bibnamefont {Hook}},\
  and\ \bibinfo {author} {\bibfnamefont {L.}~\bibnamefont {Balents}},\
  }\bibfield  {title} {\bibinfo {title} {{Topological nodal semimetals}},\
  }\href {https://doi.org/10.1103/physrevb.84.235126} {\bibfield  {journal}
  {\bibinfo  {journal} {Phys. Rev. B}\ }\textbf {\bibinfo {volume} {84}},\
  \bibinfo {pages} {235126} (\bibinfo {year} {2011})}\BibitemShut {NoStop}%
\bibitem [{\citenamefont {Armitage}\ \emph {et~al.}(2018)\citenamefont
  {Armitage}, \citenamefont {Mele},\ and\ \citenamefont
  {Vishwanath}}]{Armitage_2018}%
  \BibitemOpen
  \bibfield  {author} {\bibinfo {author} {\bibfnamefont {N.~P.}\ \bibnamefont
  {Armitage}}, \bibinfo {author} {\bibfnamefont {E.~J.}\ \bibnamefont {Mele}},\
  and\ \bibinfo {author} {\bibfnamefont {A.}~\bibnamefont {Vishwanath}},\
  }\bibfield  {title} {\bibinfo {title} {{Weyl and Dirac semimetals in
  three-dimensional solids}},\ }\href
  {https://doi.org/10.1103/revmodphys.90.015001} {\bibfield  {journal}
  {\bibinfo  {journal} {Rev. Mod. Phys.}\ }\textbf {\bibinfo {volume} {90}},\
  \bibinfo {pages} {015001} (\bibinfo {year} {2018})}\BibitemShut {NoStop}%
\bibitem [{\citenamefont {Shipunov}\ \emph {et~al.}(2020)\citenamefont
  {Shipunov}, \citenamefont {Kovalchuk}, \citenamefont {Piening}, \citenamefont
  {Labracherie}, \citenamefont {Veyrat}, \citenamefont {Wolf}, \citenamefont
  {Lubk}, \citenamefont {Subakti}, \citenamefont {Giraud}, \citenamefont
  {Dufouleur}, \citenamefont {Shokri}, \citenamefont {Caglieris}, \citenamefont
  {Hess}, \citenamefont {Efremov}, \citenamefont {B{\"u}chner},\ and\
  \citenamefont {Aswartham}}]{Shipunov_2020}%
  \BibitemOpen
  \bibfield  {author} {\bibinfo {author} {\bibfnamefont {G.}~\bibnamefont
  {Shipunov}}, \bibinfo {author} {\bibfnamefont {I.}~\bibnamefont {Kovalchuk}},
  \bibinfo {author} {\bibfnamefont {B.~R.}\ \bibnamefont {Piening}}, \bibinfo
  {author} {\bibfnamefont {V.}~\bibnamefont {Labracherie}}, \bibinfo {author}
  {\bibfnamefont {A.}~\bibnamefont {Veyrat}}, \bibinfo {author} {\bibfnamefont
  {D.}~\bibnamefont {Wolf}}, \bibinfo {author} {\bibfnamefont {A.}~\bibnamefont
  {Lubk}}, \bibinfo {author} {\bibfnamefont {S.}~\bibnamefont {Subakti}},
  \bibinfo {author} {\bibfnamefont {R.}~\bibnamefont {Giraud}}, \bibinfo
  {author} {\bibfnamefont {J.}~\bibnamefont {Dufouleur}}, \bibinfo {author}
  {\bibfnamefont {S.}~\bibnamefont {Shokri}}, \bibinfo {author} {\bibfnamefont
  {F.}~\bibnamefont {Caglieris}}, \bibinfo {author} {\bibfnamefont
  {C.}~\bibnamefont {Hess}}, \bibinfo {author} {\bibfnamefont {D.~V.}\
  \bibnamefont {Efremov}}, \bibinfo {author} {\bibfnamefont {B.}~\bibnamefont
  {B{\"u}chner}},\ and\ \bibinfo {author} {\bibfnamefont {S.}~\bibnamefont
  {Aswartham}},\ }\bibfield  {title} {\bibinfo {title} {{Polymorphic PtBi$_2$ :
  Growth, structure, and superconducting properties}},\ }\href
  {https://doi.org/10.1103/physrevmaterials.4.124202} {\bibfield  {journal}
  {\bibinfo  {journal} {Phys. Rev. Materials}\ }\textbf {\bibinfo {volume}
  {4}},\ \bibinfo {pages} {124202} (\bibinfo {year} {2020})}\BibitemShut
  {NoStop}%
\bibitem [{\citenamefont {Vocaturo}\ \emph {et~al.}(2024)\citenamefont
  {Vocaturo}, \citenamefont {Koepernik}, \citenamefont {Facio}, \citenamefont
  {Timm}, \citenamefont {Fulga}, \citenamefont {Janson},\ and\ \citenamefont
  {van~den Brink}}]{Vocaturo_2024}%
  \BibitemOpen
  \bibfield  {author} {\bibinfo {author} {\bibfnamefont {R.}~\bibnamefont
  {Vocaturo}}, \bibinfo {author} {\bibfnamefont {K.}~\bibnamefont {Koepernik}},
  \bibinfo {author} {\bibfnamefont {J.~I.}\ \bibnamefont {Facio}}, \bibinfo
  {author} {\bibfnamefont {C.}~\bibnamefont {Timm}}, \bibinfo {author}
  {\bibfnamefont {I.~C.}\ \bibnamefont {Fulga}}, \bibinfo {author}
  {\bibfnamefont {O.}~\bibnamefont {Janson}},\ and\ \bibinfo {author}
  {\bibfnamefont {J.}~\bibnamefont {van~den Brink}},\ }\bibfield  {title}
  {\bibinfo {title} {{Electronic structure of the surface-superconducting Weyl
  semimetal PtBi$_2$}},\ }\href {https://doi.org/10.1103/physrevb.110.054504}
  {\bibfield  {journal} {\bibinfo  {journal} {Phys. Rev. B}\ }\textbf {\bibinfo
  {volume} {110}},\ \bibinfo {pages} {054504} (\bibinfo {year}
  {2024})}\BibitemShut {NoStop}%
\bibitem [{\citenamefont {Mathisen}\ \emph {et~al.}(2026)\citenamefont
  {Mathisen}, \citenamefont {Tan}, \citenamefont {Brinkman}, \citenamefont
  {M{\ae}land}, \citenamefont {G{\"o}hler}, \citenamefont {Finnseth},
  \citenamefont {Shipunov}, \citenamefont {Pabst}, \citenamefont {Lemos},
  \citenamefont {Thiagarajan}, \citenamefont {Polley}, \citenamefont
  {Trauzettel}, \citenamefont {Isaeva}, \citenamefont {Facio},\ and\
  \citenamefont {Bentmann}}]{arxiv.2607.01947}%
  \BibitemOpen
  \bibfield  {author} {\bibinfo {author} {\bibfnamefont {A.~C.}\ \bibnamefont
  {Mathisen}}, \bibinfo {author} {\bibfnamefont {X.~L.}\ \bibnamefont {Tan}},
  \bibinfo {author} {\bibfnamefont {S.~S.}\ \bibnamefont {Brinkman}}, \bibinfo
  {author} {\bibfnamefont {K.}~\bibnamefont {M{\ae}land}}, \bibinfo {author}
  {\bibfnamefont {F.}~\bibnamefont {G{\"o}hler}}, \bibinfo {author}
  {\bibfnamefont {{\O}.}~\bibnamefont {Finnseth}}, \bibinfo {author}
  {\bibfnamefont {G.}~\bibnamefont {Shipunov}}, \bibinfo {author}
  {\bibfnamefont {F.}~\bibnamefont {Pabst}}, \bibinfo {author} {\bibfnamefont
  {M.~A.}\ \bibnamefont {Lemos}}, \bibinfo {author} {\bibfnamefont
  {B.}~\bibnamefont {Thiagarajan}}, \bibinfo {author} {\bibfnamefont
  {C.}~\bibnamefont {Polley}}, \bibinfo {author} {\bibfnamefont
  {B.}~\bibnamefont {Trauzettel}}, \bibinfo {author} {\bibfnamefont
  {A.}~\bibnamefont {Isaeva}}, \bibinfo {author} {\bibfnamefont {J.~I.}\
  \bibnamefont {Facio}},\ and\ \bibinfo {author} {\bibfnamefont
  {H.}~\bibnamefont {Bentmann}},\ }\bibfield  {title} {\bibinfo {title}
  {{Fermiology and spin polarization of topological surface states in
  PtBi$_2$}},\ }\href {https://doi.org/10.48550/ARXIV.2607.01947} {\bibfield
  {journal} {\bibinfo  {journal} {arXiv:2607.01947}\ } (\bibinfo {year}
  {2026})}\BibitemShut {NoStop}%
\bibitem [{\citenamefont {Crist{\'o}foli}\ \emph {et~al.}(2026)\citenamefont
  {Crist{\'o}foli}, \citenamefont {Lemos}, \citenamefont {Facio},\ and\
  \citenamefont {Cornaglia}}]{arxiv.2607.10937}%
  \BibitemOpen
  \bibfield  {author} {\bibinfo {author} {\bibfnamefont {T.}~\bibnamefont
  {Crist{\'o}foli}}, \bibinfo {author} {\bibfnamefont {M.~A.}\ \bibnamefont
  {Lemos}}, \bibinfo {author} {\bibfnamefont {J.~I.}\ \bibnamefont {Facio}},\
  and\ \bibinfo {author} {\bibfnamefont {P.~S.}\ \bibnamefont {Cornaglia}},\
  }\bibfield  {title} {\bibinfo {title} {{A minimal model for the Weyl nodes
  and Fermi arcs of PtBi$_2$}},\ }\href
  {https://doi.org/10.48550/ARXIV.2607.10937} {\bibfield  {journal} {\bibinfo
  {journal} {arXiv:2607.10937}\ } (\bibinfo {year} {2026})}\BibitemShut
  {NoStop}%
\bibitem [{\citenamefont {Brinkman}\ \emph {et~al.}(2026)\citenamefont
  {Brinkman}, \citenamefont {Tan}, \citenamefont {Mathisen}, \citenamefont
  {G{\"o}hler}, \citenamefont {Finnseth}, \citenamefont {Min}, \citenamefont
  {Shipunov}, \citenamefont {Pabst}, \citenamefont {Lemos}, \citenamefont
  {Thiagarajan}, \citenamefont {Polley}, \citenamefont {Arita}, \citenamefont
  {Shimada}, \citenamefont {Isaeva}, \citenamefont {Facio},\ and\ \citenamefont
  {Bentmann}}]{arxiv.2607.26804}%
  \BibitemOpen
  \bibfield  {author} {\bibinfo {author} {\bibfnamefont {S.~S.}\ \bibnamefont
  {Brinkman}}, \bibinfo {author} {\bibfnamefont {X.~L.}\ \bibnamefont {Tan}},
  \bibinfo {author} {\bibfnamefont {A.~C.}\ \bibnamefont {Mathisen}}, \bibinfo
  {author} {\bibfnamefont {F.}~\bibnamefont {G{\"o}hler}}, \bibinfo {author}
  {\bibfnamefont {{\O}.}~\bibnamefont {Finnseth}}, \bibinfo {author}
  {\bibfnamefont {C.-H.}\ \bibnamefont {Min}}, \bibinfo {author} {\bibfnamefont
  {G.}~\bibnamefont {Shipunov}}, \bibinfo {author} {\bibfnamefont
  {F.}~\bibnamefont {Pabst}}, \bibinfo {author} {\bibfnamefont {M.~A.}\
  \bibnamefont {Lemos}}, \bibinfo {author} {\bibfnamefont {B.}~\bibnamefont
  {Thiagarajan}}, \bibinfo {author} {\bibfnamefont {C.}~\bibnamefont {Polley}},
  \bibinfo {author} {\bibfnamefont {M.}~\bibnamefont {Arita}}, \bibinfo
  {author} {\bibfnamefont {K.}~\bibnamefont {Shimada}}, \bibinfo {author}
  {\bibfnamefont {A.}~\bibnamefont {Isaeva}}, \bibinfo {author} {\bibfnamefont
  {J.~I.}\ \bibnamefont {Facio}},\ and\ \bibinfo {author} {\bibfnamefont
  {H.}~\bibnamefont {Bentmann}},\ }\bibfield  {title} {\bibinfo {title}
  {{Disentangling bulk and surface states in the electronic structure of
  PtBi$_2$(0001)}},\ }\href {https://doi.org/10.48550/ARXIV.2607.26804}
  {\bibfield  {journal} {\bibinfo  {journal} {arXiv:2607.26804}\ } (\bibinfo
  {year} {2026})}\BibitemShut {NoStop}%
\bibitem [{\citenamefont {Zhang}\ \emph {et~al.}(2025)\citenamefont {Zhang},
  \citenamefont {Chen}, \citenamefont {Huang}, \citenamefont {Wang},
  \citenamefont {Han}, \citenamefont {Ma}, \citenamefont {Zhu}, \citenamefont
  {Ning}, \citenamefont {Shen}, \citenamefont {Huan},\ and\ \citenamefont
  {Gao}}]{Zhang_2025}%
  \BibitemOpen
  \bibfield  {author} {\bibinfo {author} {\bibfnamefont {H.}~\bibnamefont
  {Zhang}}, \bibinfo {author} {\bibfnamefont {H.}~\bibnamefont {Chen}},
  \bibinfo {author} {\bibfnamefont {Z.}~\bibnamefont {Huang}}, \bibinfo
  {author} {\bibfnamefont {Z.-A.}\ \bibnamefont {Wang}}, \bibinfo {author}
  {\bibfnamefont {G.}~\bibnamefont {Han}}, \bibinfo {author} {\bibfnamefont
  {R.}~\bibnamefont {Ma}}, \bibinfo {author} {\bibfnamefont {X.}~\bibnamefont
  {Zhu}}, \bibinfo {author} {\bibfnamefont {W.}~\bibnamefont {Ning}}, \bibinfo
  {author} {\bibfnamefont {C.}~\bibnamefont {Shen}}, \bibinfo {author}
  {\bibfnamefont {Q.}~\bibnamefont {Huan}},\ and\ \bibinfo {author}
  {\bibfnamefont {H.-J.}\ \bibnamefont {Gao}},\ }\bibfield  {title} {\bibinfo
  {title} {{Atomic Visualization of Bulk and Surface Superconductivity in Weyl
  Semimetal $\gamma$-PtBi$_2$}},\ }\href
  {https://doi.org/10.1088/0256-307x/42/12/120708} {\bibfield  {journal}
  {\bibinfo  {journal} {Chin. Phys. Lett.}\ }\textbf {\bibinfo {volume} {42}},\
  \bibinfo {pages} {120708} (\bibinfo {year} {2025})}\BibitemShut {NoStop}%
\bibitem [{\citenamefont {Stewart}(2017)}]{Stewart_2017}%
  \BibitemOpen
  \bibfield  {author} {\bibinfo {author} {\bibfnamefont {G.~R.}\ \bibnamefont
  {Stewart}},\ }\bibfield  {title} {\bibinfo {title} {{Unconventional
  superconductivity}},\ }\href {https://doi.org/10.1080/00018732.2017.1331615}
  {\bibfield  {journal} {\bibinfo  {journal} {Adv. Phys.}\ }\textbf {\bibinfo
  {volume} {66}},\ \bibinfo {pages} {75} (\bibinfo {year} {2017})}\BibitemShut
  {NoStop}%
\bibitem [{\citenamefont {Buccheri}\ \emph {et~al.}(2026)\citenamefont
  {Buccheri}, \citenamefont {de~Martino},\ and\ \citenamefont {van~den
  Brink}}]{Buccheri_2026}%
  \BibitemOpen
  \bibfield  {author} {\bibinfo {author} {\bibfnamefont {F.}~\bibnamefont
  {Buccheri}}, \bibinfo {author} {\bibfnamefont {A.}~\bibnamefont
  {de~Martino}},\ and\ \bibinfo {author} {\bibfnamefont {J.}~\bibnamefont
  {van~den Brink}},\ }\bibfield  {title} {\bibinfo {title} {{Phonon-driven
  nodal surface superconductivity of Fermi arcs}},\ }\href
  {https://doi.org/10.48550/ARXIV.2606.02371} {\bibfield  {journal} {\bibinfo
  {journal} {arXiv:2606.02371}\ } (\bibinfo {year} {2026})}\BibitemShut
  {NoStop}%
\bibitem [{\citenamefont {M{\ae}land}\ \emph {et~al.}(2026)\citenamefont
  {M{\ae}land}, \citenamefont {Sangiovanni},\ and\ \citenamefont
  {Trauzettel}}]{Maeland_2026}%
  \BibitemOpen
  \bibfield  {author} {\bibinfo {author} {\bibfnamefont {K.}~\bibnamefont
  {M{\ae}land}}, \bibinfo {author} {\bibfnamefont {G.}~\bibnamefont
  {Sangiovanni}},\ and\ \bibinfo {author} {\bibfnamefont {B.}~\bibnamefont
  {Trauzettel}},\ }\bibfield  {title} {\bibinfo {title} {{Mechanism for Nodal
  Topological Superconductivity on PtBi$_2$ Surface}},\ }\href
  {https://doi.org/10.1103/x1cy-w5zd} {\bibfield  {journal} {\bibinfo
  {journal} {Phys. Rev. Lett.}\ }\textbf {\bibinfo {volume} {137}},\ \bibinfo
  {pages} {056001} (\bibinfo {year} {2026})}\BibitemShut {NoStop}%
\bibitem [{\citenamefont {Dsouza}\ \emph {et~al.}(2026)\citenamefont {Dsouza},
  \citenamefont {Parthenios}, \citenamefont {Andersen},\ and\ \citenamefont
  {Christensen}}]{Dsouza_2026}%
  \BibitemOpen
  \bibfield  {author} {\bibinfo {author} {\bibfnamefont {R.}~\bibnamefont
  {Dsouza}}, \bibinfo {author} {\bibfnamefont {N.}~\bibnamefont {Parthenios}},
  \bibinfo {author} {\bibfnamefont {B.~M.}\ \bibnamefont {Andersen}},\ and\
  \bibinfo {author} {\bibfnamefont {M.~H.}\ \bibnamefont {Christensen}},\
  }\bibfield  {title} {\bibinfo {title} {{Kohn-Luttinger Superconductivity of
  Weyl Fermi Arcs in PtBi$_2$}},\ }\href
  {https://doi.org/10.48550/ARXIV.2605.31501} {\bibfield  {journal} {\bibinfo
  {journal} {arXiv:2605.31501}\ } (\bibinfo {year} {2026})}\BibitemShut
  {NoStop}%
\bibitem [{\citenamefont {Waje}\ \emph {et~al.}(2025)\citenamefont {Waje},
  \citenamefont {Jakubczyk}, \citenamefont {van~den Brink},\ and\ \citenamefont
  {Timm}}]{Waje_2025}%
  \BibitemOpen
  \bibfield  {author} {\bibinfo {author} {\bibfnamefont {H.}~\bibnamefont
  {Waje}}, \bibinfo {author} {\bibfnamefont {F.}~\bibnamefont {Jakubczyk}},
  \bibinfo {author} {\bibfnamefont {J.}~\bibnamefont {van~den Brink}},\ and\
  \bibinfo {author} {\bibfnamefont {C.}~\bibnamefont {Timm}},\ }\bibfield
  {title} {\bibinfo {title} {{Ginzburg-Landau theory for unconventional surface
  superconductivity in PtBi$_2$}},\ }\href {https://doi.org/10.1103/kkqg-ntcz}
  {\bibfield  {journal} {\bibinfo  {journal} {Phys. Rev. B}\ }\textbf {\bibinfo
  {volume} {112}},\ \bibinfo {pages} {144519} (\bibinfo {year}
  {2025})}\BibitemShut {NoStop}%
\bibitem [{\citenamefont {Semenoff}(1984)}]{Semenoff_1984}%
  \BibitemOpen
  \bibfield  {author} {\bibinfo {author} {\bibfnamefont {G.~W.}\ \bibnamefont
  {Semenoff}},\ }\bibfield  {title} {\bibinfo {title} {{Condensed-Matter
  Simulation of a Three-Dimensional Anomaly}},\ }\href
  {https://doi.org/10.1103/physrevlett.53.2449} {\bibfield  {journal} {\bibinfo
   {journal} {Phys. Rev. Lett.}\ }\textbf {\bibinfo {volume} {53}},\ \bibinfo
  {pages} {2449} (\bibinfo {year} {1984})}\BibitemShut {NoStop}%
\bibitem [{\citenamefont {Haldane}(1988)}]{Haldane_1988}%
  \BibitemOpen
  \bibfield  {author} {\bibinfo {author} {\bibfnamefont {F.~D.~M.}\
  \bibnamefont {Haldane}},\ }\bibfield  {title} {\bibinfo {title} {{Model for a
  Quantum Hall Effect without Landau Levels: Condensed-Matter Realization of
  the “Parity Anomaly”}},\ }\href
  {https://doi.org/10.1103/physrevlett.61.2015} {\bibfield  {journal} {\bibinfo
   {journal} {Phys. Rev. Lett.}\ }\textbf {\bibinfo {volume} {61}},\ \bibinfo
  {pages} {2015} (\bibinfo {year} {1988})}\BibitemShut {NoStop}%
\bibitem [{\citenamefont {Men’shov}\ \emph {et~al.}(2013)\citenamefont
  {Men’shov}, \citenamefont {Tugushev}, \citenamefont {Eremeev},
  \citenamefont {Echenique},\ and\ \citenamefont {Chulkov}}]{Men_shov_2013}%
  \BibitemOpen
  \bibfield  {author} {\bibinfo {author} {\bibfnamefont {V.~N.}\ \bibnamefont
  {Men’shov}}, \bibinfo {author} {\bibfnamefont {V.~V.}\ \bibnamefont
  {Tugushev}}, \bibinfo {author} {\bibfnamefont {S.~V.}\ \bibnamefont
  {Eremeev}}, \bibinfo {author} {\bibfnamefont {P.~M.}\ \bibnamefont
  {Echenique}},\ and\ \bibinfo {author} {\bibfnamefont {E.~V.}\ \bibnamefont
  {Chulkov}},\ }\bibfield  {title} {\bibinfo {title} {{Magnetic proximity
  effect in the three-dimensional topological insulator/ferromagnetic insulator
  heterostructure}},\ }\href {https://doi.org/10.1103/physrevb.88.224401}
  {\bibfield  {journal} {\bibinfo  {journal} {Phys. Rev. B}\ }\textbf {\bibinfo
  {volume} {88}},\ \bibinfo {pages} {224401} (\bibinfo {year}
  {2013})}\BibitemShut {NoStop}%
\bibitem [{\citenamefont {Chang}\ \emph {et~al.}(2013)\citenamefont {Chang},
  \citenamefont {Zhang}, \citenamefont {Feng}, \citenamefont {Shen},
  \citenamefont {Zhang}, \citenamefont {Guo}, \citenamefont {Li}, \citenamefont
  {Ou}, \citenamefont {Wei}, \citenamefont {Wang}, \citenamefont {Ji},
  \citenamefont {Feng}, \citenamefont {Ji}, \citenamefont {Chen}, \citenamefont
  {Jia}, \citenamefont {Dai}, \citenamefont {Fang}, \citenamefont {Zhang},
  \citenamefont {He}, \citenamefont {Wang}, \citenamefont {Lu}, \citenamefont
  {Ma},\ and\ \citenamefont {Xue}}]{Chang_2013}%
  \BibitemOpen
  \bibfield  {author} {\bibinfo {author} {\bibfnamefont {C.-Z.}\ \bibnamefont
  {Chang}}, \bibinfo {author} {\bibfnamefont {J.}~\bibnamefont {Zhang}},
  \bibinfo {author} {\bibfnamefont {X.}~\bibnamefont {Feng}}, \bibinfo {author}
  {\bibfnamefont {J.}~\bibnamefont {Shen}}, \bibinfo {author} {\bibfnamefont
  {Z.}~\bibnamefont {Zhang}}, \bibinfo {author} {\bibfnamefont
  {M.}~\bibnamefont {Guo}}, \bibinfo {author} {\bibfnamefont {K.}~\bibnamefont
  {Li}}, \bibinfo {author} {\bibfnamefont {Y.}~\bibnamefont {Ou}}, \bibinfo
  {author} {\bibfnamefont {P.}~\bibnamefont {Wei}}, \bibinfo {author}
  {\bibfnamefont {L.-L.}\ \bibnamefont {Wang}}, \bibinfo {author}
  {\bibfnamefont {Z.-Q.}\ \bibnamefont {Ji}}, \bibinfo {author} {\bibfnamefont
  {Y.}~\bibnamefont {Feng}}, \bibinfo {author} {\bibfnamefont {S.}~\bibnamefont
  {Ji}}, \bibinfo {author} {\bibfnamefont {X.}~\bibnamefont {Chen}}, \bibinfo
  {author} {\bibfnamefont {J.}~\bibnamefont {Jia}}, \bibinfo {author}
  {\bibfnamefont {X.}~\bibnamefont {Dai}}, \bibinfo {author} {\bibfnamefont
  {Z.}~\bibnamefont {Fang}}, \bibinfo {author} {\bibfnamefont {S.-C.}\
  \bibnamefont {Zhang}}, \bibinfo {author} {\bibfnamefont {K.}~\bibnamefont
  {He}}, \bibinfo {author} {\bibfnamefont {Y.}~\bibnamefont {Wang}}, \bibinfo
  {author} {\bibfnamefont {L.}~\bibnamefont {Lu}}, \bibinfo {author}
  {\bibfnamefont {X.-C.}\ \bibnamefont {Ma}},\ and\ \bibinfo {author}
  {\bibfnamefont {Q.-K.}\ \bibnamefont {Xue}},\ }\bibfield  {title} {\bibinfo
  {title} {{Experimental Observation of the Quantum Anomalous Hall Effect in a
  Magnetic Topological Insulator}},\ }\href
  {https://doi.org/10.1126/science.1234414} {\bibfield  {journal} {\bibinfo
  {journal} {Science}\ }\textbf {\bibinfo {volume} {340}},\ \bibinfo {pages}
  {167} (\bibinfo {year} {2013})}\BibitemShut {NoStop}%
\bibitem [{\citenamefont {Read}\ and\ \citenamefont {Green}(2000)}]{Read_2000}%
  \BibitemOpen
  \bibfield  {author} {\bibinfo {author} {\bibfnamefont {N.}~\bibnamefont
  {Read}}\ and\ \bibinfo {author} {\bibfnamefont {D.}~\bibnamefont {Green}},\
  }\bibfield  {title} {\bibinfo {title} {{Paired states of fermions in two
  dimensions with breaking of parity and time-reversal symmetries and the
  fractional quantum Hall effect}},\ }\href
  {https://doi.org/10.1103/physrevb.61.10267} {\bibfield  {journal} {\bibinfo
  {journal} {Phys. Rev. B}\ }\textbf {\bibinfo {volume} {61}},\ \bibinfo
  {pages} {10267} (\bibinfo {year} {2000})}\BibitemShut {NoStop}%
\bibitem [{\citenamefont {Timm}\ and\ \citenamefont
  {Bhattacharya}(2021)}]{Timm_2021}%
  \BibitemOpen
  \bibfield  {author} {\bibinfo {author} {\bibfnamefont {C.}~\bibnamefont
  {Timm}}\ and\ \bibinfo {author} {\bibfnamefont {A.}~\bibnamefont
  {Bhattacharya}},\ }\bibfield  {title} {\bibinfo {title} {{Symmetry, nodal
  structure, and Bogoliubov Fermi surfaces for nonlocal pairing}},\ }\href
  {https://doi.org/10.1103/physrevb.104.094529} {\bibfield  {journal} {\bibinfo
   {journal} {Phys. Rev. B}\ }\textbf {\bibinfo {volume} {104}},\ \bibinfo
  {pages} {094529} (\bibinfo {year} {2021})}\BibitemShut {NoStop}%
\bibitem [{\citenamefont {Groth}\ \emph {et~al.}(2014)\citenamefont {Groth},
  \citenamefont {Wimmer}, \citenamefont {Akhmerov},\ and\ \citenamefont
  {Waintal}}]{Groth_2014}%
  \BibitemOpen
  \bibfield  {author} {\bibinfo {author} {\bibfnamefont {C.~W.}\ \bibnamefont
  {Groth}}, \bibinfo {author} {\bibfnamefont {M.}~\bibnamefont {Wimmer}},
  \bibinfo {author} {\bibfnamefont {A.~R.}\ \bibnamefont {Akhmerov}},\ and\
  \bibinfo {author} {\bibfnamefont {X.}~\bibnamefont {Waintal}},\ }\bibfield
  {title} {\bibinfo {title} {{Kwant: a software package for quantum
  transport}},\ }\href {https://doi.org/10.1088/1367-2630/16/6/063065}
  {\bibfield  {journal} {\bibinfo  {journal} {New J. Phys.}\ }\textbf {\bibinfo
  {volume} {16}},\ \bibinfo {pages} {063065} (\bibinfo {year}
  {2014})}\BibitemShut {NoStop}%
\bibitem [{\citenamefont {Wu}\ \emph {et~al.}(2020)\citenamefont {Wu},
  \citenamefont {Sun}, \citenamefont {Zhao},\ and\ \citenamefont
  {Pan}}]{Wu_2020}%
  \BibitemOpen
  \bibfield  {author} {\bibinfo {author} {\bibfnamefont {M.-M.}\ \bibnamefont
  {Wu}}, \bibinfo {author} {\bibfnamefont {Y.}~\bibnamefont {Sun}}, \bibinfo
  {author} {\bibfnamefont {H.}~\bibnamefont {Zhao}},\ and\ \bibinfo {author}
  {\bibfnamefont {H.}~\bibnamefont {Pan}},\ }\bibfield  {title} {\bibinfo
  {title} {{Quantum confinement effect induced topological phase transitions in
  anisotropic Weyl semimetal}},\ }\href
  {https://doi.org/10.1016/j.spmi.2019.106386} {\bibfield  {journal} {\bibinfo
  {journal} {Superlattices Microstruct.}\ }\textbf {\bibinfo {volume} {138}},\
  \bibinfo {pages} {106386} (\bibinfo {year} {2020})}\BibitemShut {NoStop}%
\bibitem [{\citenamefont {Altland}\ and\ \citenamefont
  {Zirnbauer}(1997)}]{Altland_1997}%
  \BibitemOpen
  \bibfield  {author} {\bibinfo {author} {\bibfnamefont {A.}~\bibnamefont
  {Altland}}\ and\ \bibinfo {author} {\bibfnamefont {M.~R.}\ \bibnamefont
  {Zirnbauer}},\ }\bibfield  {title} {\bibinfo {title} {{Nonstandard symmetry
  classes in mesoscopic normal-superconducting hybrid structures}},\ }\href
  {https://doi.org/10.1103/physrevb.55.1142} {\bibfield  {journal} {\bibinfo
  {journal} {Phys. Rev. B}\ }\textbf {\bibinfo {volume} {55}},\ \bibinfo
  {pages} {1142} (\bibinfo {year} {1997})}\BibitemShut {NoStop}%
\bibitem [{\citenamefont {Avron}\ \emph {et~al.}(1983)\citenamefont {Avron},
  \citenamefont {Seiler},\ and\ \citenamefont {Simon}}]{Avron_1983}%
  \BibitemOpen
  \bibfield  {author} {\bibinfo {author} {\bibfnamefont {J.~E.}\ \bibnamefont
  {Avron}}, \bibinfo {author} {\bibfnamefont {R.}~\bibnamefont {Seiler}},\ and\
  \bibinfo {author} {\bibfnamefont {B.}~\bibnamefont {Simon}},\ }\bibfield
  {title} {\bibinfo {title} {{Homotopy and Quantization in Condensed Matter
  Physics}},\ }\href {https://doi.org/10.1103/physrevlett.51.51} {\bibfield
  {journal} {\bibinfo  {journal} {Phys. Rev. Lett.}\ }\textbf {\bibinfo
  {volume} {51}},\ \bibinfo {pages} {51} (\bibinfo {year} {1983})}\BibitemShut
  {NoStop}%
\bibitem [{\citenamefont {Bellissard}\ \emph {et~al.}(1994)\citenamefont
  {Bellissard}, \citenamefont {van Elst},\ and\ \citenamefont
  {Schulz-Baldes}}]{Bellissard_1994}%
  \BibitemOpen
  \bibfield  {author} {\bibinfo {author} {\bibfnamefont {J.}~\bibnamefont
  {Bellissard}}, \bibinfo {author} {\bibfnamefont {A.}~\bibnamefont {van
  Elst}},\ and\ \bibinfo {author} {\bibfnamefont {H.}~\bibnamefont
  {Schulz-Baldes}},\ }\bibfield  {title} {\bibinfo {title} {{The noncommutative
  geometry of the quantum Hall effect}},\ }\href
  {https://doi.org/10.1063/1.530758} {\bibfield  {journal} {\bibinfo  {journal}
  {J. Math. Phys.}\ }\textbf {\bibinfo {volume} {35}},\ \bibinfo {pages} {5373}
  (\bibinfo {year} {1994})}\BibitemShut {NoStop}%
\bibitem [{\citenamefont {Fukui}\ \emph {et~al.}(2005)\citenamefont {Fukui},
  \citenamefont {Hatsugai},\ and\ \citenamefont {Suzuki}}]{Fukui_2005}%
  \BibitemOpen
  \bibfield  {author} {\bibinfo {author} {\bibfnamefont {T.}~\bibnamefont
  {Fukui}}, \bibinfo {author} {\bibfnamefont {Y.}~\bibnamefont {Hatsugai}},\
  and\ \bibinfo {author} {\bibfnamefont {H.}~\bibnamefont {Suzuki}},\
  }\bibfield  {title} {\bibinfo {title} {{Chern Numbers in Discretized
  Brillouin Zone: Efficient Method of Computing (Spin) Hall Conductances}},\
  }\href {https://doi.org/10.1143/jpsj.74.1674} {\bibfield  {journal} {\bibinfo
   {journal} {J. Phys. Soc. Jpn.}\ }\textbf {\bibinfo {volume} {74}},\ \bibinfo
  {pages} {1674} (\bibinfo {year} {2005})}\BibitemShut {NoStop}%
\bibitem [{\citenamefont {Palacio-Morales}\ \emph {et~al.}(2019)\citenamefont
  {Palacio-Morales}, \citenamefont {Mascot}, \citenamefont {Cocklin},
  \citenamefont {Kim}, \citenamefont {Rachel}, \citenamefont {Morr},\ and\
  \citenamefont {Wiesendanger}}]{Palacio_Morales_2019}%
  \BibitemOpen
  \bibfield  {author} {\bibinfo {author} {\bibfnamefont {A.}~\bibnamefont
  {Palacio-Morales}}, \bibinfo {author} {\bibfnamefont {E.}~\bibnamefont
  {Mascot}}, \bibinfo {author} {\bibfnamefont {S.}~\bibnamefont {Cocklin}},
  \bibinfo {author} {\bibfnamefont {H.}~\bibnamefont {Kim}}, \bibinfo {author}
  {\bibfnamefont {S.}~\bibnamefont {Rachel}}, \bibinfo {author} {\bibfnamefont
  {D.~K.}\ \bibnamefont {Morr}},\ and\ \bibinfo {author} {\bibfnamefont
  {R.}~\bibnamefont {Wiesendanger}},\ }\bibfield  {title} {\bibinfo {title}
  {{Atomic-scale interface engineering of Majorana edge modes in a 2D
  magnet-superconductor hybrid system}},\ }\href
  {https://doi.org/10.1126/sciadv.aav6600} {\bibfield  {journal} {\bibinfo
  {journal} {Sci. Adv.}\ }\textbf {\bibinfo {volume} {5}},\ \bibinfo {pages}
  {eaav6600} (\bibinfo {year} {2019})}\BibitemShut {NoStop}%
\bibitem [{\citenamefont {Jiao}\ \emph {et~al.}(2020)\citenamefont {Jiao},
  \citenamefont {Howard}, \citenamefont {Ran}, \citenamefont {Wang},
  \citenamefont {Rodriguez}, \citenamefont {Sigrist}, \citenamefont {Wang},
  \citenamefont {Butch},\ and\ \citenamefont {Madhavan}}]{Jiao_2020}%
  \BibitemOpen
  \bibfield  {author} {\bibinfo {author} {\bibfnamefont {L.}~\bibnamefont
  {Jiao}}, \bibinfo {author} {\bibfnamefont {S.}~\bibnamefont {Howard}},
  \bibinfo {author} {\bibfnamefont {S.}~\bibnamefont {Ran}}, \bibinfo {author}
  {\bibfnamefont {Z.}~\bibnamefont {Wang}}, \bibinfo {author} {\bibfnamefont
  {J.~O.}\ \bibnamefont {Rodriguez}}, \bibinfo {author} {\bibfnamefont
  {M.}~\bibnamefont {Sigrist}}, \bibinfo {author} {\bibfnamefont
  {Z.}~\bibnamefont {Wang}}, \bibinfo {author} {\bibfnamefont {N.~P.}\
  \bibnamefont {Butch}},\ and\ \bibinfo {author} {\bibfnamefont
  {V.}~\bibnamefont {Madhavan}},\ }\bibfield  {title} {\bibinfo {title}
  {{Chiral superconductivity in heavy-fermion metal UTe$_2$}},\ }\href
  {https://doi.org/10.1038/s41586-020-2122-2} {\bibfield  {journal} {\bibinfo
  {journal} {Nature}\ }\textbf {\bibinfo {volume} {579}},\ \bibinfo {pages}
  {523} (\bibinfo {year} {2020})}\BibitemShut {NoStop}%
\bibitem [{\citenamefont {Ketmaier}\ \emph {et~al.}(2026)\citenamefont
  {Ketmaier}, \citenamefont {van~den Brink},\ and\ \citenamefont
  {Fulga}}]{Zenodo_code}%
  \BibitemOpen
  \bibfield  {author} {\bibinfo {author} {\bibfnamefont {L.}~\bibnamefont
  {Ketmaier}}, \bibinfo {author} {\bibfnamefont {J.}~\bibnamefont {van~den
  Brink}},\ and\ \bibinfo {author} {\bibfnamefont {I.~C.}\ \bibnamefont
  {Fulga}},\ }\bibfield  {title} {\bibinfo {title} {{Tunable Chern
  superconductivity of PtBi$_2$ in slab geometry}},\ }\href
  {https://doi.org/10.5281/zenodo.22279094} {\bibfield  {journal} {\bibinfo
  {journal} {https://doi.org/10.5281/zenodo.22279094}\ } (\bibinfo {year}
  {2026})}\BibitemShut {NoStop}%
\end{thebibliography}%

\end{document}